\documentclass{aastex62}

\usepackage{natbib}
\usepackage{amsmath}
\usepackage{url}

\newcommand{\myemail}{billylquarles@gmail.com}

\shorttitle{Kepler-62f Obliquity Evolution}
\shortauthors{Quarles, Barnes, Lissauer, \& Chambers}

\begin{document}

\title{OBLIQUITY EVOLUTION OF THE POTENTIALLY HABITABLE EXOPLANET KEPLER-62F}

\author{Billy Quarles}
\affil{HL Dodge Department of Physics \& Astronomy, University of Oklahoma, Norman, OK 73019,
USA}
		\email{\myemail}
\author{Jason W. Barnes}
\affil{Department of Physics, University of Idaho, Moscow, Idaho 83844-0903, USA}
\author{Jack J. Lissauer}
\affil{NASA Ames Research Center, Astrobiology and Space Science Division MS 245-3,
    Moffett Field, CA 94035, USA}
\author{John Chambers}
\affil{Department of Terrestrial Magnetism, Carnegie Institution for Science, Washington, DC 20015, USA}


\begin{abstract}
Variations in the axial tilt, or obliquity, of terrestrial planets can affect their climates and therefore their habitability.  Kepler-62f is a 1.4 R$_\oplus$ planet orbiting within the habitable zone of its K2 dwarf host star \citep{Borucki2013}.  We perform N-body simulations that monitor the evolution of obliquity of Kepler-62f for 10 million year timescales to explore the effects on model assumptions, such as the masses of the Kepler-62 planets and the possibility of outer bodies.  Significant obliquity variation occurs when the rotational precession frequency overlaps with one or more of the secular orbital frequencies, but most variations are limited to $\lesssim$10$^\circ$.  Moderate variations ($\sim$10$^\circ - 20^\circ$) can occur over a broader range of initial obliquities when the relative nodal longitude ($\Delta\Omega$) overlaps with the frequency and phase of a given secular mode.  However, we find that adding outer gas giants on long period orbits ($\gtrsim$ 1000 days) can produce large ($\sim$60$^\circ$) variations in obliquity if Kepler-62f has a very rapid (4 hr) rotation period.  {The possibility of giant planets on long period orbits impacts the climate and habitability of Kepler-62f through variations in the latitudinal surface flux, where the timescale for large variation can occur on million year timescales.}   
 
\end{abstract}


  \section{Introduction}
The {\it Kepler} mission has discovered more than 1000 extrasolar planets \citep[e.g.,][]{Lissauer2014,Rowe2014} that represent a broad spectrum of possible worlds. The {\it Kepler} planets represent a reservoir of outcomes from planet formation.  From this, researchers can explore and test our models of how planets behave.  In this paper, we focus on the evolution of a planet's axial tilt, or obliquity, which can be modified by the neighboring planets within a system through the gravitational torque that they exert on its equatorial bulge.  

\citeauthor{Borucki2013} (2013; henceforth BAF13) announced the discovery of a five planet system orbiting a K2 dwarf, which is a star slightly smaller in mass and radius than our Sun.  Interestingly, the outer two planets in this system (Kepler-62e \& Kepler-62f) may reside within a region of space where liquid water could exist on their surfaces {given a rocky surface and an atmosphere that permits a reasonable greenhouse effect}, otherwise known as the habitable zone.  This star, Kepler-62, and its planets, Kepler-62b -- Kepler-62f, occupy a region of parameter space that is exciting to explore and can have astrobiological implications.

Among the parameters that govern a planet's astrobiological potential  is the obliquity, or axis tilt, $\psi$.  Our Earth with a middling $\psi_\oplus=23.44^\circ$ enjoys moderate seasonal  weather variability \citep{Williams1997,Spiegel2009}.  If a planet's obliquity gets \emph{too} low, then a lack of illumination at high latitudes can lead to polar glaciations.  Planets with  obliquities above $\psi\approx56^\circ$ would experience severe seasons, alternately baking and freezing their poles while their equators receive the least annual illumination on the planet.  That high-obliquity regime applies to both the present-day Pluto \citep{White2017,Howard2017} and Mars in the geologic past \citep{Mege2011,Head2014}.   

{The study of obliquity evolution of the solar system planets has a long history, where \cite{Ward1973} showed that the obliquity of Mars undergoes large variations due to the perturbations of the planets.  Later, \cite{Touma1993} showed that these variations are in fact chaotic, and \cite{Laskar1993a} {performed} a frequency analysis to the remaining terrestrial planets of the Solar System.  \cite{Laskar1993b} have shown that the obliquity of the Earth would be also chaotic in absence of the Moon over a range of initial rotation periods and obliquities.  \cite{Correia2003a} and \cite{Correia2003b} have shown that the past obliquity of Venus was chaotic and may have aided the planet to transition between a prograde and retrograde rotation.  \cite{Ward2004} and \cite{Hamilton2004} used the planetary perturbations to explain the present high obliquity of Saturn (through a resonance with Neptune).  The planetary perturbations were also proposed as the mechanism to tilt Uranus during the initial migration stages in the solar system \citep{Brunini2006}, although this scenario was later discarded \citep{Lee2007}.  \cite{Brasser2011} showed that the Martian obliquity during the Noachian era could be outside the chaotic regime.}

\citeauthor{Lissauer2012} (2012; henceforth LBC12) showed that the obliquity of a hypothetical Earth in the absence of a large Moon typically varies by $\sim\pm10^\circ$, in contrast to (albeit not in conflict with) prior calculations showing an allowed range of  $0^\circ<\psi_\oplus<85^\circ$ \citep{Laskar1993a}.  {The difference between these studies is two-fold: (1) LBC12 employed a full n-body method where the prior calculations were based on a secular solution and (2) LBC12 evolved the Solar System for $\pm$2 Gyr where the timescale to explore the full region requires a longer timescale.  Recently, \cite{Saillenfest2019} have updated the procedure for using the secular solutions and have highlighted possible applications to exoplanets.}

\citet{Barnes2016} analyzed the possible obliquity variations of early Venus in part as a possible analog for habitable exoplanets, but direct studies of exoplanet obliquity variations have been lacking.  Recently, \cite{Deitrick2018a} demonstrated semi-analytical methods for evaluating obliquity evolution and the connection to exo-Milankovitch cycles \citep{Spiegel2010}.  \cite{Shields2016} performed a limited dynamical study of Kepler-62f starting the interior planets from circular orbits while allowing Kepler-62f to begin with a moderate eccentricity.  \cite{Shields2016} used their dynamical model in 3D climate simulations to estimate potential climates for the planet.  However, these results depend on the model assumptions assumed in the dynamical simulations, where the orbital architecture of the system is largely incomplete.

In this paper, we investigate the potential obliquity variability of Kepler-62f under a range of model assumptions, including the possible existence of outer bodies and two different mass-radius relations to determine the planetary masses.  The outer bodies we consider are analogs of the gas giants in the solar system
so that our results are comparable with previous studies in terms of dynamics
and habitability.  Our numerical method builds upon the work of LBC12 and is described in Section 2, where our model assumptions concerning the initial orbital architectures of our realizations are also presented.  Section 3 details the results of our simulations and discusses interesting facets from within the possible outcomes and the possible impact on habitability due to flux variations.  Section 4 presents the general conclusions that we may draw from this study.

\section{Methodology}

\subsection{Estimating the Masses of the Observed Planets}

Many parameters within Kepler-62 are well characterized by BAF13, including the orbital periods and planetary radii, but they were not able to determine masses for any of the five planets.  They performed an analysis focused on transit timing measurements within the data that yielded upper limits on the masses, but these upper limits do not exclude any physically plausible compositions.  \cite{Bolmont2015} have performed calculations for Kepler-62 and they make the assumption that all five planets are rocky in order to prescribe masses.  They give caution to their study especially considering the effects of tides and rotational flattening that are sensitive to the assumed composition.  In this work, we assume the smallest 3 or 4 of the planets to be rocky and the largest planet(s) to be more massive but less dense than the rocky planets.

Besides the deficit of knowledge of mass in the known planetary bodies, there is also an incomplete knowledge of the full architecture of this system.  The {\it Kepler} mission provided nearly four years of continuous monitoring, but other planetary bodies may exist in the Kepler-62 system with larger orbital periods.  Moreover, planets can also escape detection through a small misalignment of inclination relative to our line of sight.  We thus perform many possible realizations in the rotational state of Kepler-62f with and without giant planets on long period orbits in order to identify the sensitivity of the obliquity evolution on our model assumptions.

Following \cite{Lissauer2013}, we estimate the masses of the planets using a mass-radius relation in the form of a piecewise power law, $M_p/M_\oplus = \xi (R_p/R_\oplus)^\epsilon$, where $M_p$ is the mass of the planet, $R_p$ is the radius of the planet, $\xi$ is a scale factor, and $\epsilon$ is the power.  This power law is divided into three regimes as follows:


\begin{equation}
\label{MR_relation}
\frac{M_p}{M_\oplus} = \xi \left(\frac{R_p}{R_\oplus}\right)^\epsilon 
\begin{cases} 
      \xi = 1, \epsilon = \frac{1}{0.3}; & R_p < 1\: R_\oplus \\ 
      \xi = 1, \epsilon = \frac{1}{0.27}; & 1\: R_\oplus \leq R_p \leq \beta \: R_\oplus \\
      \xi = \beta^{\frac{1}{0.27} - 1}, \epsilon = 1;  & R_p> \beta \: R_\oplus ,\\ 
   \end{cases}
\end{equation}

\noindent where the last regime uses a parameter $\beta$ that defines the assumed transition point between ``rocky'' and  volatile-rich planets and an assumed slope in the volatile-rich regime that is within the range of those derived from fits to measured values for exoplanets \citep{Weiss2014,Wolfgang2016,Chen2017}.

The obliquity evolution of a planet is only of astrobiological interest if it is rocky.  Planets in the habitable zone with $R_p = 1.41\: R_\oplus$ may well be rocky, while those with $R_p = 1.61\: R_\oplus$ are probably not \citep{Rogers2015}.  For most of our simulations, we assume $\beta = 1.41$ (Case A) where we consider Kepler-62e to be volatile rich and $\beta=1.61$ (Case B) that allows Kepler-62e to be rocky.  Table \ref{tab:Mass} shows the nominal values in the planet radius from BAF13 and the masses for each planet for each assumed composition of Kepler-62e.  Kepler-62d is considered to be volatile-rich in both cases, but using a different transition approximately doubles its mass.  However, this will likely only modify slightly the perturbations on Kepler-62f because its semimajor axis is $\sim$5x greater.

\begin{deluxetable}{lccccc}
\tablecolumns{6}
\tablewidth{0pc}
\tablecaption{Assumed Radii and Masses for the Kepler-62 Planets\label{tab:Mass}}
\tablehead{\colhead{} & \colhead{62b} & \colhead{62c} & \colhead{62d} & \colhead{62e} & \colhead{62f}} 
\startdata
$R_p$ (R$_\oplus$) & 1.31 & 0.54 & 1.95 & 1.61 & 1.41 \\
$M_p$ (M$_\oplus$), Case A  & 2.71 & 0.128 & 4.94 & 4.05 & 3.57 \\
$M_p$ (M$_\oplus$), Case B & 2.71 & 0.128 & 7.06 & 5.83 & 3.57 \\
\enddata
\tablecomments{Assumed masses for the planets considering Kepler-62e to be either volatile-rich (Case A) or ``rocky'' (Case B).}

\end{deluxetable}

\subsection{Orbital Solution from Observations}
BAF13 provided orbital parameters derived from a statistical analysis using a transit model to compare with the observational data.  Using the results of BAF13, we define or derive all the orbital elements to uniquely prescribe a starting orbit and list those parameters in Table \ref{tab:ICs}.  Using the observed orbital period ratios, we prescribe for each planet a semimajor axis ($a$) assuming that the mass of the host star is 0.69 M$_\odot$.  We also use the prescribed masses in Table \ref{tab:Mass} for this calculation, but their inclusion are largely negligible and augment the semimajor axes of the outer two planets by $\sim$10$^{-4}$ AU.

The results of BAF13 provide the components of the eccentricity vector ($e \cos \omega, e\sin \omega$), which we convert to the dynamicist convention through a sign change in $\omega$.  BAF13 also acknowledge that the nominal values provided result in an unstable configuration leading to the ejection of Kepler-62c, the Mars-sized planet.  Most of our planets are assigned an eccentricity using the nominal values of the eccentricity vectors given in BAF13.  But, we start our simulations choosing the $1\sigma$ upper bound value of $e\sin \omega$ component for Kepler-62c (--~0.07 rather than the nominal value of --~0.18) and use the nominal values for $e\cos \omega$.  By choosing the eccentricity of Kepler-62c in this way, we found that the orbits in the Kepler-62 system to be dynamically stable up to 100 Myr.     


\begin{deluxetable}{lcccccccc}
\rotate
\tablecolumns{6}
\tablewidth{0pc}
\tablecaption{Initial Orbital Parameters Used for the Kepler-62 Planets\label{tab:ICs}}
\tablehead{\colhead{Planet} & \colhead{Period (days)} & \colhead{$a$ (AU)} & \colhead{$e$} & \colhead{$i$ (deg.)} & \colhead{$\omega$ (deg.)} & \colhead{$\Omega$ (deg.)} & \colhead{$M$ (deg.)} & \colhead{$T_{eq}$ (K)}} 
\startdata
Kepler-62b & 5.714932 & 0.0552802   &	0.07071068  & 0.883 &	 278.13010  &	90.373788  & 74.589661 & 750 \\
Kepler-62c & 12.4417 & 0.0928576	  & 0.08602325	& 0.560 &    54.462322	& 90.473841	 & 327.80383  & 578 \\
Kepler-62d & 18.16406 & 0.1195013	  & 0.09486833	& 0.300 &   288.43494	& 90.002197	 & 243.09602 & 510 \\
Kepler-62e & 122.3874 & 0.4263096	  & 0.13000000  & 0.303 &	 292.61986	& 90.303058  & 63.220181 &  270 \\
Kepler-62f & 267.291 & 0.7176171 	& 0.09433981	& 0.488 &	  57.994616	& 90.478155	 & 176.54639 & 208
\enddata
\tablecomments{Initial values of the orbital elements used in our simulations for the Kepler-62 system.  The radiative equilibrium temperature, $T_{eq}$, is also provided to delineate each planet relative to its potential habitability.  See BAF13 for the true uncertainties in these parameters.}

\end{deluxetable}

 Also in the dynamicist convention, we measure orbital inclination relative to the line of sight (90$^\circ$ from the sky plane) and define the ascending node ($\Omega$) to be 90$^\circ$ during the time of transit.  The observations tell us that all five planets will have nearly identical values of ascending node by the virtue that they all transit.  But we don't know precisely what the values are and we assume them to be within a degree of the node defining the line of sight.  {By analyzing the ratios of transit durations for planets in Kepler's multi-planet systems, statistical studies show that the typical mutual inclination of planets within Kepler multiplanet systems are $<$2.2$^\circ$ \citep{Fang2012,Fabrycky2014,Ballard2016,Moriarty2016}}.  To define this numerically, we add randomly generated values between 0$^\circ$ -- 0.5$^\circ$ to the line of sight value (90$^\circ$).  After defining the ascending node for each planet, we determine the orbital inclination relative to the line of sight through the spherical law of cosines.  From the parameters defined thus far, we determine the mean anomaly ($M$) of each planet at the epoch of central transit as given by the data and then find the mean anomalies relative to a common epoch.  

\subsection{Possible Outer Bodies} \label{sec:outer_bodies}
Our best current methods for indirectly detecting exoplanets is inherently biased towards bodies with relatively short orbital periods, but more planets may exist at longer periods \citep{Mills2019}.  The existence of these bodies could have an astrobiological impact on exoplanets within the habitable zone without rendering the overall system unstable.  In order to determine the impact on obliquity, we include 2 sets of analogs to the giant planets, where one is similar to those in the Solar System (SS) and another is drawn from a large number of stability simulations of 2 planet pairs of Jupiter- and Saturn-mass planets.  These analogs are identical in mass to their Solar System giant namesakes.  The set that follows the orbital elements of the Solar System giants \citep[See appendix A][]{Murray2000} are scaled in semimajor axis by their orbital period (see Table \ref{tab:ICs_gas}) due to the less massive host star.

{The orbital period of our Jupiter-analog ($\sim$4300 days) is quite large and its perturbations on the inner system do not affect their stability.  Also, a set of giant planets at such distances may not produce the largest changes to the spin evolution of Kepler-62f.  In order to identify the most extreme conditions for the obliquity evolution of Kepler-62f we need to identify plausible orbital elements for gas giants where the invoked bodies could have escaped detection and the inner system remains stable.}  Thus, we perform a suite of $\sim$15,000 simulations of using a pair of gas giants (Jupiter- and Saturn-mass) over a range of orbital periods and evaluate whether the system can remain stable for 10 Myr given random initial conditions for the pair of gas giants.  In these simulations, we use the masses of the inner planets from Case A.  

The initial conditions for the Jupiter mass planet are drawn from a uniform distribution in period ranging from 300 -- 1600 days, Rayleigh distributions in the eccentricity ($\sigma_e = 0.05$) and inclination ($\sigma_i = 1^\circ$), a uniform distribution in the ascending node ranging from 85$^\circ$ -- 95$^\circ$, and uniform distributions for the argument of periastron and mean anomaly ranging from 0$^\circ$ -- 360$^\circ$.  The initial conditions for the Saturn mass planet are chosen in a similar fashion for most of the orbital elements.  Instead of drawing from a distribution in period, we use a uniform distribution ranging from 10 -- 20 R$_H$, where R$_H = a_J(M_J/(3M_*))^{1 / 3}$ represents the Hill Radius, to ensure the stability of the pair of gas giants \citep{Gladman1993,Chambers1996}.  

From these simulations, we find that systems are stable when we choose a Jupiter analog with an orbital period $\gtrsim$1000 days with a corresponding Saturn analog outside of mean motion resonance.  Thus, we choose 1 Jupiter-Saturn pair to include in our exploration of obliquity with a Jupiter analog (orbital period $\sim$1084 days) and a Saturn-analog separated by $\sim$11 R$_H$, which is wide of the 5:2 mean motion resonance.  By choosing our setup in this manner, we will have a giant planet architecture that will substantially perturb the inner system without causing a global instability and provide a much larger perturbation to the possible obliquity of Kepler-62f.    {Employing a hypothetical pair of giant planets is important because it provides a broader context to the study of habitability of Kepler-62f through an investigation of very extreme conditions.}  {The initial orbital elements are given in Table \ref{tab:ICs_gas} for the gas giants drawn from the Solar System (SS) and our randomly drawn gas giant pair (RG).} 

\begin{deluxetable}{lccccccc}
\tablecolumns{6}
\tablewidth{0pc}
\tablecaption{Initial Orbital Parameters Used For Our Gas Giants\label{tab:ICs_gas}}
\tablehead{\colhead{Planet} & \colhead{Period (days)} & \colhead{$a$ (AU)} & \colhead{$e$} & \colhead{$i$ (deg.)} & \colhead{$\omega$ (deg.)} & \colhead{$\Omega$ (deg.)} & \colhead{$M$ (deg.)}} 
\startdata
SS$_J$ & 4336.1086 & 4.598637392  & 0.04839266  & 1.3053  &	 -85.8023  &	100.55615  & 19.65053  \\
SS$_S$ & 10757.994 & 8.42784611  & 0.0541506	  & 2.48446 &  -21.2831	& 113.71504	 & -42.48762  \\
SS$_U$ & 30707.225 & 16.95857856 & 0.04716771  & 0.76986 &   96.73436	& 74.22988	 & 142.26794 \\
SS$_N$ & 60223.18 & 26.57081264 & 0.00858587  & 1.76917 &	 -86.75034	& 131.72169  & 259.90868  \\
\hline
RG$_J$ & 1083.6600 & 1.82455841 	& 0.03328494	& 0.8819402 &	  51.9209548	& 93.8755285	 & 210.832984 \\
RG$_S$ & 2781.5082 & 3.42044515 	& 0.03193422	& 2.286905  &	  234.538802	& 89.5462986	 & 351.061437 
\enddata
\tablecomments{Initial values of the orbital elements used in our simulations using the gas giants of the Solar System (SS) and a selected stable configuration of randomly drawn gas giant pairs (RG).  The subscripts denote the mass of each planet by the respective analog within the Solar System (Jupiter -- Neptune).}

\end{deluxetable}

\subsection{Numerical Setup for Obliquity Evolution}
To evaluate the obliquity evolution, we use a modified version of the \texttt{smercury} integration package (LBC12) that has been optimized for determining the extrema in obliquity evolution up to a given integration step a{nd uses the formalism developed in \cite{Touma1993}}.  We define the obliquity as the mutual inclination, through the spherical law of cosines, between the spin axis ($i_s$, $\Omega_s$) and orbital axis ($i$, $\Omega$).  The nodal difference, $\Delta \Omega = \Omega_s - \Omega$, is set to 0$^\circ$ in a majority of our simulations, where a subset of our 5 planet systems using the Case B masses begin with $\Delta\Omega = 90^\circ$.

The short orbital period of Kepler-62b ($\sim$5.715 days) poses a numerical challenge for evaluating a broad and deep range of parameters.  As a result, we limit our simulations to 10 Myr using a timestep (0.286 days) that is 5\% of the orbital period of Kepler-62b.  One avenue that could be employed to reach longer times is to remove the inner two planets (Kepler-62b and Kepler-62c), where they are added to the mass of the host star effectively increasing the $J_2$ of the host star.  This would allow a larger timestep to be chosen relative to the orbital period of Kepler-62d.  However, much of our discussion on the variation of obliquity depends on the secular frequencies of the system and removing the inner planets would shift the relevant frequencies.  {\cite{Farago2009} used the averaged Hamiltonian of an inner planet in order to evaluate the orbital evolution of more distant planets on longer timescales, but this method is beyond the scope of our current study.}  Our numerical code, \texttt{smercury}, does not include possible tidal interactions and thus keeps the rotation period constant throughout the simulation.  This is justified because our simulations do not reach the timescales ($\sim$ 1 Gyr) necessary for tides to be important for Kepler-62f.

{Previous works \citep{Laskar1993a,Li2014a,Li2014b,Shan2018,Deitrick2018a} have used the secular solution for obliquity, while more recent studies have used N-body methods that include spin-orbit interactions \citep{Lissauer2012,Bolmont2015,Barnes2016}.  However, we can make comparisons to historical formalisms that use secular solutions through relevant precession frequencies.}  \cite{Laskar1993a}, \cite{Shan2018}, and others use the 'precession' constant $\alpha$ measured in arcseconds per year, which is defined as follows:

\begin{equation} \label{eq:alpha}
\alpha = {3 n^2 \over 2 \nu}{C-A \over C} \approx {3n^2 \over 2 \nu} {J_2 \over \bar{C}},
\end{equation}

where $n$ represents the mean motion, $\nu$ denotes the rotational frequency, $\bar{C}$ relates to the moment of Inertia, and $J_2$ is the zonal harmonic related to the flattening due the rotation.  The moment of Inertia and $J_2$ are presently unknown, so we use the values assumed in LBC12, where $J_2$ is derived using the Darwin-Radau Relation \citep[see Appendix A][]{Lissauer2012}.

Figure \ref{fig:spin_param} shows how our assumptions on the planet masses in Case A relate to the equatorial radius (R$_{eq}$), the derived zonal harmonic (J$_2$), and the approximate value of the precession constant, $\alpha$, and Table \ref{tab:rot_params} provides specific values for select rotation periods.  We note that Figure \ref{fig:spin_param} (bottom panel) shows the precession constant for Kepler-62e (red dashed line) to very high ($>$ 60 \arcsec/yr), even for slow rotation periods (P$_{rot} >$ 40 hr).  As a result, the obliquity variations of the planet will likely be small for the rotational parameters that we consider and instead focus on the possible obliquity variations of Kepler-62f. 

\begin{deluxetable}{cccc}
\tablecolumns{6}
\tablewidth{0pc}
\tablecaption{Initial Rotational Parameters Used in Our Simulations\label{tab:rot_params}}
\tablehead{\colhead{P$_{rot}$} & \colhead{R$_{eq}$} & \colhead{J$_2$} & \colhead{$\alpha$} \\
\colhead{(hr)} & \colhead{(km)} & & \colhead{(\arcsec/yr)}} 
\startdata
4	&	9316.35	&	0.0331066	&	166.81803	 \\
6	&	9116.36	&	0.0137867	&	104.20265	 \\
8	&	9055.86	&	0.0076016	&	76.60628	 \\
10	&	9029.19	&	0.0048222	&	60.74512	 \\
12	&	9015.03	&	0.0033330	&	50.38312	 \\
14	&	9006.60	&	0.0024419	&	43.06447	 \\
16	&	9001.17	&	0.0018662	&	37.61329	 \\
18	&	8997.46	&	0.0014727	&	33.39277	 \\
20	&	8994.82	&	0.0011918	&	30.02705	 \\
22	&	8992.87	&	0.0009843	&	27.27958	 \\
24	&	8991.39	&	0.0008267	&	24.99394	 \\
26	&	8990.24	&	0.0007041	&	23.06249	 \\
28	&	8989.33	&	0.0006070	&	21.40866	 \\
30	&	8988.60	&	0.0005286	&	19.97652	 \\
32	&	8988.00	&	0.0004645	&	18.72423	 \\
34	&	8987.50	&	0.0004114	&	17.61988	 \\
36	&	8987.08	&	0.0003669	&	16.63868	 \\
38	&	8986.73	&	0.0003293	&	15.76111	 \\
40	&	8986.43	&	0.0002971	&	14.97155	 \\
42	&	8986.17	&	0.0002695	&	14.25739	 \\
44	&	8985.95	&	0.0002455	&	13.60831	 \\
46	&	8985.75	&	0.0002246	&	13.01579	 \\
48	&	8985.58	&	0.0002063	&	12.47275	 
\enddata
\tablecomments{Initial values for rotation period (P$_{rot}$), equatorial radius (R$_{eq}$), zonal harmonic (J$_2$), and precession constant ($\alpha$) for Kepler-62f determined from LBC12.  These values are the same for both Case A and Case B as the mass of Kepler-62f is unchanged between the two cases.}

\end{deluxetable}

Significant variation of retrograde obliquities ($\psi_o >90^\circ$) takes a long-time to develop computationally and dynamically, even in under less computationally demanding conditions \citep{Barnes2016}, so this work will focus mainly on the obliquity evolution of prograde ($\psi_o \leq 90^\circ$) rotators.  Another unknown parameter is the rotation period of the planets, where we explore a wide range (4 -- 24 hours) for the runs considering the 5 planet system, Kepler-62b -- Kepler-62f.  We extend this range to 48 hours for the cases including the outer bodies because longer orbital periods can introduce lower frequencies that can potentially overlap.  By taking steps in the rotation period rather than precession frequency \citep[e.g.,][]{Laskar1993a,Laskar1993b}, we seek to identify the larger structures with the parameter space.

\subsection{Calculating the Surface Flux}
The potential habitability of an exoplanet is hard to define and often depends on a range of assumed parameters that influence the exchange of energy between the subsurface, atmospheric, and local space environment \citep{Kasting1993,Kopparapu2014,Kopparapu2016,Ramirez2018}.  Therefore, our study is limited in terms of the energy received at the top layer of the atmosphere, or surface, of Kepler-62f.  The atmospheric composition and albedo of Kepler-62f, which are unknown, are necessary to provide realistic estimates of the temperature
variations on the surface of the planet. We consider as a proxy for potential habitability.  For this, we consider the surface flux, $S_p$ as a function of the stellar luminosity ($L_\star$ in $L_\odot$), the instantaneous stellar distance ($r$ in AU), and the solar constant ($S_\oplus$) to get:
\begin{align}
\label{eq:flux}
S_p &= \frac{L_\star}{L_\odot} \left(\frac{1 AU}{r}\right)^2 S_\oplus.
\end{align}  
The instantaneous stellar distance can be obtained through numerical integration of an orbit, but the variation of orbital parameters (semimajor axis and eccentricity) is small on the timescale of a single orbit, so that it can be computed over all values of the true anomaly, $f$, by
\begin{align}
r &= \frac{a(1-e^2)}{1+e\cos f}.
\end{align}

However, the surface flux at any point on the planet will vary as a function of latitude on a sphere.  To incorporate this effect, we define the daily mean top-of-atmosphere {insolation}, $I_d$ in $W/m^2$, at any point as,
\begin{align}
I_d &= \frac{S_p}{\pi}\left[\eta \sin \delta_\star \sin \delta + \sin(\eta) \cos \delta_\star \cos \delta \right],
\end{align} 
where $\eta$ is the half-angle of daylight (i.e., a measure of the day length in radians) at a given latitude, $\delta$, and the substellar latitude, $\delta_\star = \psi \cos(f + \Delta f)$, that is determined by the obliquity ($\psi$), the true anomaly ($f$), and an orbital phase offset ($\Delta f$), or the offset in orbital phase between periastron and the highest solar declination in the northern hemisphere \citep[e.g.,][]{Armstrong2004,Armstrong2014,Shields2016,Kane2017}.  The orbital phase offset for any exoplanet is unknown, where we use $\Delta f=0.25$ throughout our work.  An orbital phase of zero corresponds to the planet's periastron passage \citep[e.g.,][]{Kane2017}.  The half-angle of daylight is computed through the following conditions:
\begin{equation}
\label{eq:eta}
\cos \eta = 
\begin{cases} 
-\tan \delta \tan \delta_\star; & |\delta| < 90^\circ - |\delta_\star| \\ 
-1; & \delta - \delta_\star \leq -90^\circ \: {\rm or} \: \delta - \delta_\star \leq 90^\circ\\  
1; & \delta + \delta_\star \geq 90^\circ \: {\rm or} \: \delta + \delta_\star \leq 90^\circ\\ 
\end{cases}
\end{equation}
following \cite{Armstrong2014}.  Using Equations \ref{eq:flux} - \ref{eq:eta}, we calculate the latitudinal flux incident on Kepler-62f and average over the orbital phase to identify how the flux changes annually as a function of latitude.

\section{Results}
Our simulations investigate the obliquity evolution of Kepler-62f for a 10 Myr timescale.  We examine two different assumptions on the mass, and thereby compositions, of the 5 known (transiting) planets, Case A and Case B. Planets on long-period orbits have a low transit probability.  Therefore, hypothetical outer planets are also included in many of our simulations.  In some cases, we add to the transiting planets a set of giant planet analogs drawn from the Solar System (SS) and scaled by period are also included.  In others, randomly determined giant planets (RG) are also used with initial conditions drawn from the results of our stability simulations (see Section \ref{sec:outer_bodies}).  

\begin{figure}
\centering
\epsscale{1.0}
\includegraphics[width=\linewidth]{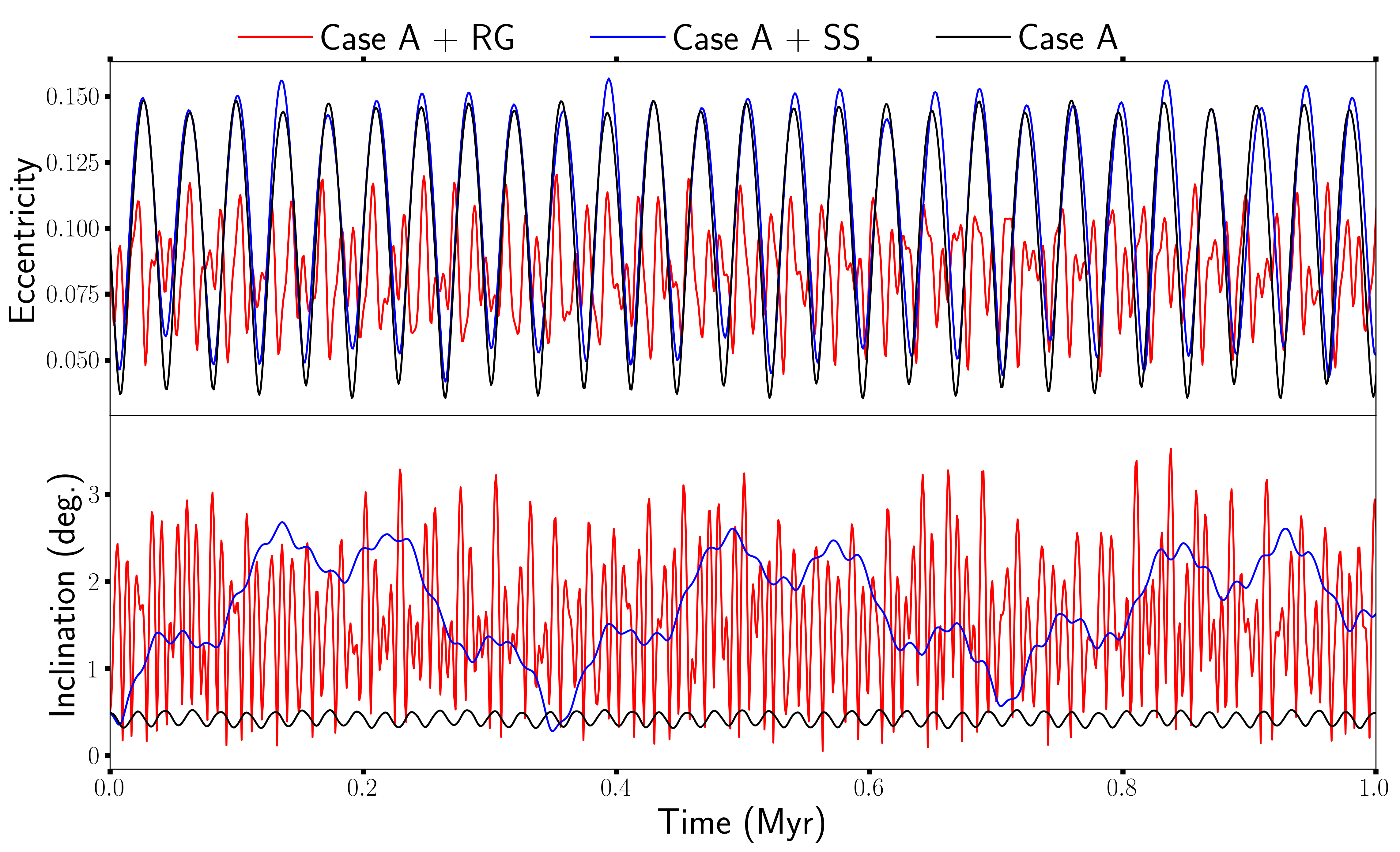}
\caption{Evolution of the eccentricity (top) and inclination (bottom) of Kepler-62f using the planet masses from Case A (black).  The evolution of these parameters including the scaled Solar System giants (SS, blue) and our random gas giant pair (RG, red) are also shown.}
\label{fig:K62f_orb}
\end{figure}

The inclusion of giant planets alters the eccentricity and inclination of Kepler-62f over time.  Figure \ref{fig:K62f_orb} shows these effects using the 5 planet system (black), the 9 planet system with the scaled Solar System giants (blue), and the 7 planet system with the randomly drawn giant planet pair (red).  Much of our analysis relates to overlap between axial precession frequencies and the secular orbital frequencies.  Sections \ref{sec:5pl_var}, and \ref{sec:outer_var} identifies \emph{where} certain frequencies are important in relation to our initial rotational states and Section \ref{sec:freq_overlap} focuses more on \emph{why} those frequencies (and other factors) are important.  Given our broad range of rotation periods, the range in the respective precession constant will also be large and our figures account for this by using a base-10 logarithmic scale for the precession constant.  We highlight the areas where select frequencies may overlap by black curves in Figs. \ref{fig:CaseA}, \ref{fig:CaseB}, \ref{fig:CaseA_SS}, and \ref{fig:CaseA_RG} using the $f_j$ values given in Table \ref{tab:freq}.

\subsection{Variations Due to the Transiting Planets} \label{sec:5pl_var}
The Kepler-62 system consists of 5 transiting planets, all of whose orbital periods are shorter than the Earth's, and 3 of the planets have periods shorter than Mercury's.  As a result the orbits are relatively close together and can produce perturbations on neighboring planets that in turn influence the evolution of obliquity.  \cite{Bolmont2015},  using a tidal model and incorporating General Relativity effects over Gyr timescales, showed that the rotation periods of the planets interior to Kepler-62f can be substantially slowed leading to a state near $\psi = 0^\circ$.  However, the changes to Kepler-62f under the same model experiences much smaller changes to its obliquity and rotation period.  Thus, we examine a \emph{broad} range of initial rotation periods (4 -- 24 hr) for a prograde Kepler-62f using both of our assumptions for the masses of the planets (Case A and Case B) on a 10 Myr timescale.

\begin{figure}
\centering
\epsscale{1.0}
\includegraphics[width=\linewidth]{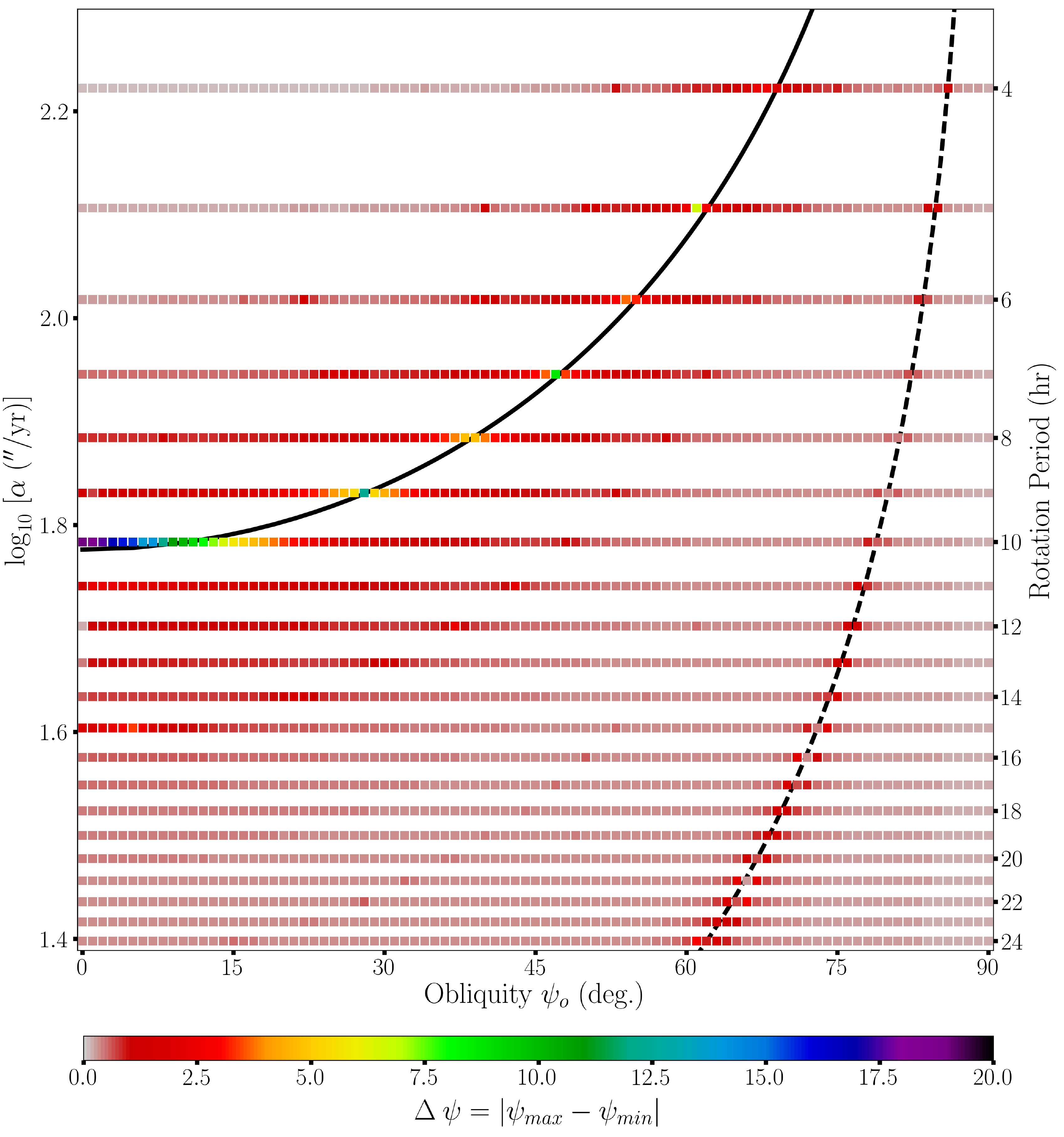}
\caption{Obliquity variations of Kepler-62f using the 5 planet system assuming a mass-radius transition at 1.41 $R_\oplus$ (Case A).  The color scale denotes the range in obliquity variation, $|\psi_{max} - \psi_{min}|$, obtained for each simulation over a 10 Myr timescale.  The left vertical axis marks the value of the precession constant, $\alpha$, on a logarithmic scale, where the right vertical axis provides the corresponding rotation period (see eq. \ref{eq:alpha}).  The black curves (solid \& dashed) represent where the respective orbital precession frequencies ($f_2$ \& $f_3$) in Table \ref{tab:freq} overlap with the precession constant.}
\label{fig:CaseA}
\end{figure}

Figure \ref{fig:CaseA} illustrates the results of these simulations for nominal values of the planetary masses (Case A). By using the range of obliquity variation, $\Delta \psi \equiv |\psi_{max} - \psi_{min}|$, we estimate the most likely values of obliquity variation and how they depend on the initial rotation state of Kepler-62f.  The most common $\Delta\psi$ values in Fig. \ref{fig:CaseA} are less than $\sim$3$^\circ$, which is roughly similar to the amount of variation for the present-day Earth with the Moon ($\sim$2.4$^\circ$).  There are distinct regions where $\Delta\psi \lesssim$ 1$^\circ$ (light gray) is more prevalent than those $> 1^\circ$, but $< 3^\circ$ (red).

The largest $\Delta\psi$ values occur at $\psi_o = 0^\circ$ with a 10 hr rotation period (purple) and corresponds to a precession constant $\sim$60 \arcsec/yr (see Table \ref{tab:rot_params}).  This amounts to an instantaneous precession period of $\sim$22,000 years, which is less than the present-day Earth's.  For the 10 hr rotation period, $\Delta\psi$ decreases as the initial obliquity, $\psi_o$, increases until $\sim$16$^\circ$.  The formula for the expected axial precession, $\dot{\Omega}_s$, is
\begin{equation} \label{eq:axial}
\dot{\Omega}_s =-f_j= \alpha\cos\:\psi,
\end{equation}
where $f_j$ denotes the modal frequency (see Sec. \ref{sec:freq_overlap}).  Using Eq. \ref{eq:axial}, the precession frequency is $\sim$62.5 \arcsec/yr for $\psi = 16^\circ$ and $\alpha \approx 60$ \arcsec/yr.  Thus, the decline in obliquity variation depends on the proximity to the expected precession frequency.  We show the evolution of obliquity in a representation similar to a phase portrait in Figure \ref{fig:K62f_phase}, where a fixed point appears $\psi\sim$16$^\circ$.  When Kepler-62f begins with a 9 hr rotation period ($\alpha \sim 67.7$ \arcsec/yr), the highest variation appears at $\psi_o = 28^\circ$, where using Equation \ref{eq:axial} we find $\dot{\Omega}_s \approx 60$ \arcsec/yr and the width around this peak in $\Delta\psi$ is smaller.  This trend continues for faster (decreasing) rotation periods over the range that we simulated.

Figure \ref{fig:CaseB} demonstrates a similar exploration, but considers a different set of masses for the two largest transiting planets, Kepler-62d and Kepler-62e (Case B).  Since the mass of Kepler-62f remains unchanged, the values for the zonal harmonic, J$_2$, and the precession constant, $\alpha$, also remain unchanged.  Naively, we would expect to see the same result as in Fig. \ref{fig:CaseA}.  The overall features in Fig. \ref{fig:CaseB} are similar to Fig. \ref{fig:CaseA}, but the largest variation appears at a shorter rotation period (8 hr), where we would expect --73 \arcsec/yr to be the significant frequency due to the larger masses.  This is likely caused by an induced precession from the increased mass for Kepler-62e (see Section \ref{sec:freq_overlap}).  Kepler-62d nearly doubles in mass between the two cases, but it is much farther away from Kepler-62f than is Kepler-62e ($\sim$40\% mass increase in Case B).  {Both Figs. \ref{fig:CaseA} and \ref{fig:CaseB} differ from the solar system due to the compactness of the system.  There is more variation in the similar plots of the solar system \citep{Laskar1993a} due to the slower orbital precession of the outer giant overlapping with plausible spin precessions of the inner planets.} 

\subsection{Effects of Outer Giant Planets} \label{sec:outer_var}
The full architecture of the Kepler-62 systems (i.e., number of planets and masses) is largely unknown, so  outer perturbers could introduce other precession frequencies.  We explore three scenarios where outer giant planets may exist: (1) a scaled version of the Solar System giant planets, (2) the scaled Solar System giant planets with their orbital inclinations doubled, and (3) a Jupiter-Saturn pair of planets drawn from stability simulations (see Section \ref{sec:outer_bodies}).  All three scenarios are performed using the masses from Case A, but only the first scenario is performed using the masses from Case B.  Also, we include rotation periods beyond 24 hr to show the expected variations at lower frequencies ($\sim$20 \arcsec/yr).  The precession frequencies when adding the Solar System giant planets differ from those in the Solar System because we did not scale masses of the giant planets and the relative semimajor axis between Kepler-62f and the giant planets is substantially different \citep[][Chap. 7]{Murray2000}.

Including a scaled version of the Solar System giant planets (using the Case A masses) introduces more variation, $\Delta\psi$, at higher initial obliquities and longer rotation periods.  Figure \ref{fig:CaseA_SS} demonstrates this result; note that the color scale has been adjusted in response.  The most common values of $\Delta\psi$ are $\sim$3$^\circ$ -- 5$^\circ$ (red), which is a little higher than in Figure \ref{fig:CaseA}.  The location of the largest variation also changes in response to the perturbations of the outer planets on the orbit of Kepler-62f.  The regions of large variation ($\Delta\psi > 5^\circ$) typically occupy regions of the parameter space with a short rotation period or high ($>45^\circ$) initial obliquity.  However, there are regions with larger variations for rotation periods longer than 24 hr.  Figure \ref{fig:CaseB_SS} (using the Case B masses) shows similar differences compared to Fig. \ref{fig:CaseB} with larger variations in the same regions of parameter space.  Similar to Fig. \ref{fig:CaseA_SS} the most common variation in Fig. \ref{fig:CaseB_SS} increases to $\sim$5$^\circ$.

In the second and third scenarios, the differences in large scale structure is the most interesting, where we use the masses from Case A.  Thus, we evaluate only the even rotation periods for the full range from 4 -- 48 hr.  Doubling the inclination of the Solar System giant planets increases the overall variation, as shown in Figure \ref{fig:CaseA_SS_2x}, where the largest variation increases to $\sim$42$^\circ$.  This occurs in the low initial obliquity range, but for slower rotation periods ($>$36 hr).  {Part of this increase can be attributed to the larger range of values possible when the orbital angular momentum vector decouples from the spin vector ($\Delta\psi \sim 2i$).}  One may expect that doubling the inclinations would also increase the potential for large variations in retrograde ($\psi_o >90^\circ$).  We perform a set of runs exploring initially retrograde obliquities for this scenario and find that for $\psi_o \geq 95^\circ$, the obliquity varies up to $5^\circ$, which is similar to our prograde ($\psi_o \leq 90 ^\circ$) results (i.e., red points in Fig. \ref{fig:CaseA_SS_2x}).  When $\psi_o$  starts in the 91$^\circ$ -- $94^\circ$ range, more substantial variations ($\Delta\psi \sim$26$^\circ$) can occur, but they quickly decrease with increasing initial obliquity.  In the third scenario, the lowest variation in obliquity is $\sim$3$^\circ$ and the most common value is $\sim$10$^\circ$, as shown in Figure \ref{fig:CaseA_RG}.  For a very fast rotator (P$_{rot}$ = 4 hr), the obliquity variation can be quite large, up to $\sim$65$^\circ$ even on the relatively short timescale (10 Myr) of our simulations.   

\subsection{Effects From Overlapping Frequencies} \label{sec:freq_overlap}
\begin{figure}
\centering
\epsscale{1.0}
\includegraphics[width=\linewidth]{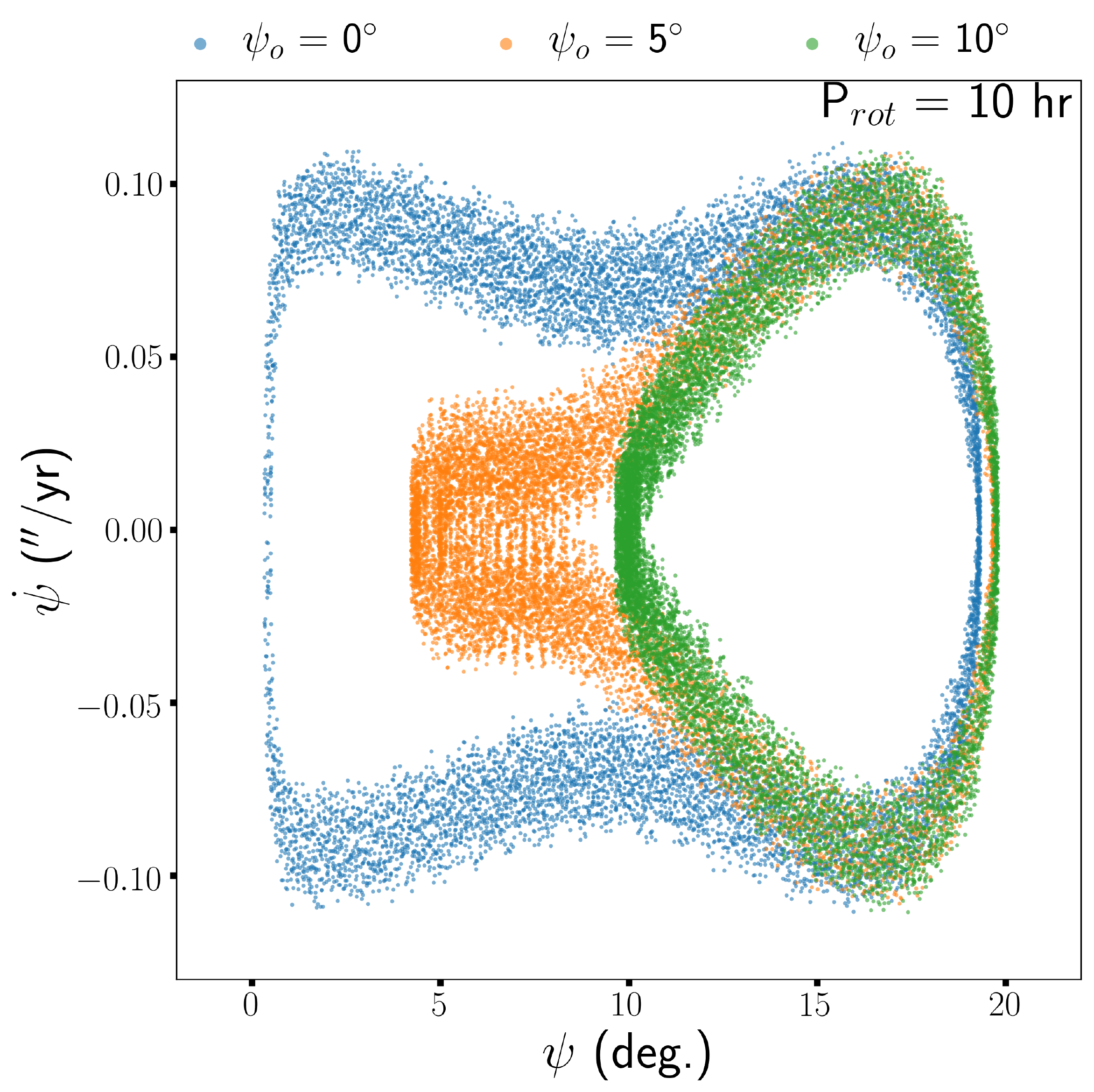}
\caption{Phase-like portrait of the obliquity ($\psi$ in deg.) and the numerical derivative ($\dot{\psi}$ in \arcsec/yr) for three initial values of obliquity ($\psi_o = 0^\circ - 10^\circ$) assuming the masses from Case A and a rotation period of 10 hr.}
\label{fig:K62f_phase}
\end{figure}

\begin{figure}
\centering
\epsscale{1.0}
\includegraphics[width=\linewidth]{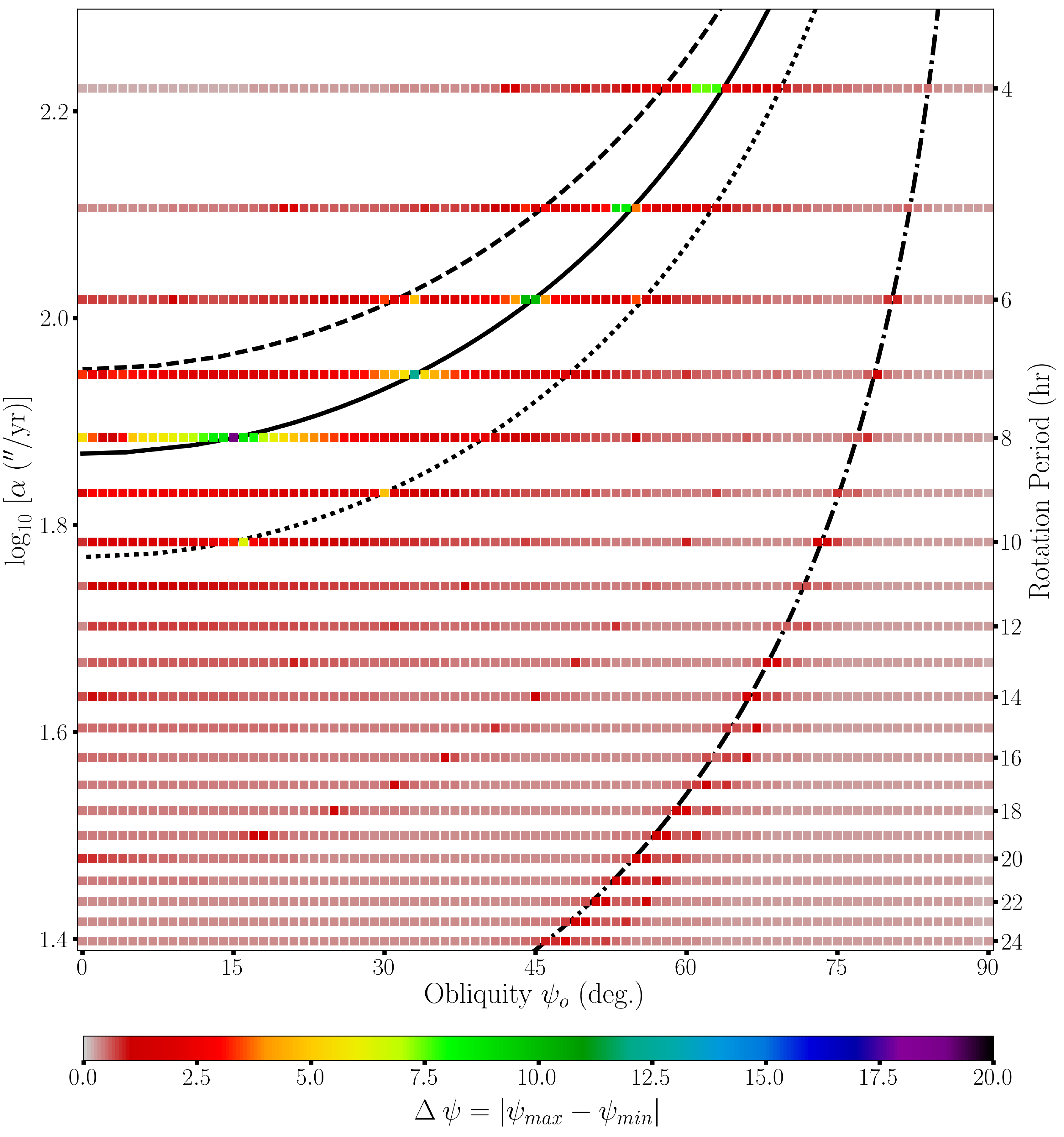}
\caption{Similar to Figure \ref{fig:CaseA}, but assuming a mass-radius transition at 1.61 $R_\oplus$ (Case B).  The black curves (solid, dashed, dotted, \& dash-dot) represent where the respective orbital precession frequencies ($f_2$, $f_3$, $f_4$, \& $f_5$) in Table \ref{tab:freq} overlap with the precession constant.}
\label{fig:CaseB}
\end{figure}

\begin{figure}
\centering
\epsscale{1.0}
\includegraphics[width=\linewidth]{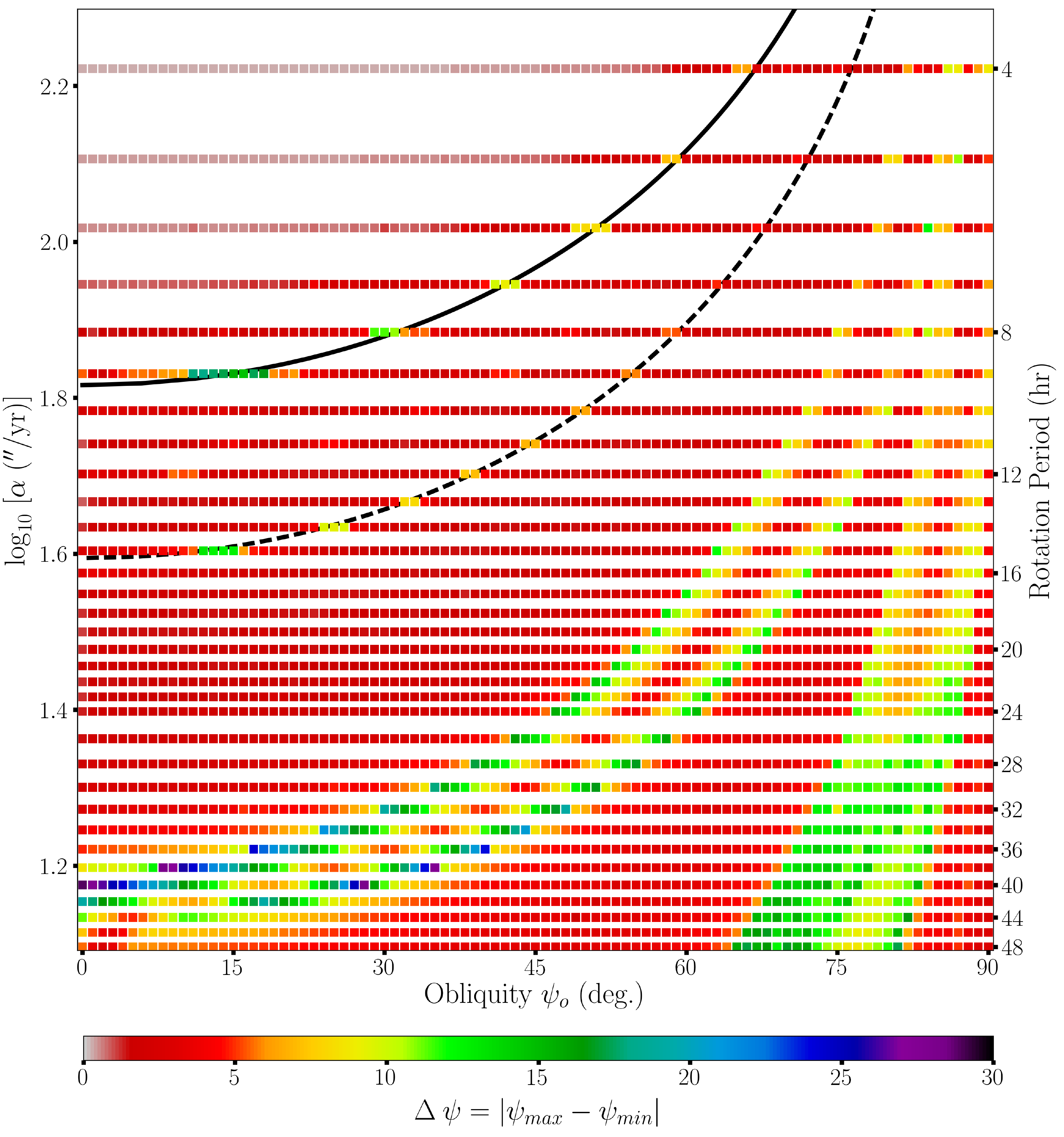}
\caption{Similar to Figure \ref{fig:CaseA} with the addition of 4 giant planets similar to the Solar System in mass and orbital architecture but scaled by period (Case A + SS).  Even rotation periods beyond 24 hr are included to demonstrate the additional variation due to the giant planets at smaller frequencies.  As a result, the maximum value of the color scale is changed. The black curves (solid \& dashed) represent where the respective orbital precession frequencies ($f_6$ \& $f_7$) in Table \ref{tab:freq} overlap with the precession constant.}
\label{fig:CaseA_SS}
\end{figure}

\begin{figure}
\centering
\epsscale{1.0}
\includegraphics[width=\linewidth]{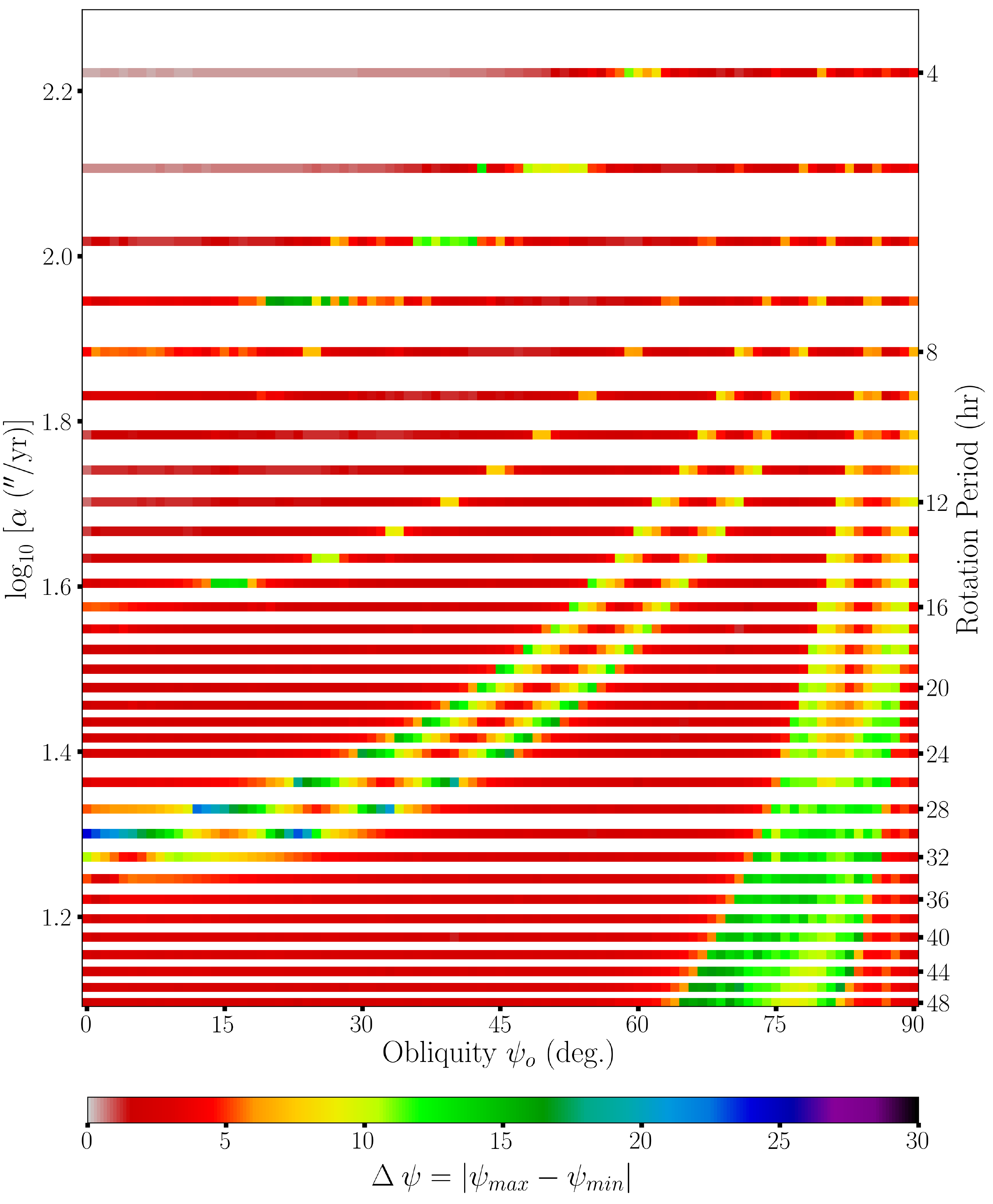}
\caption{Similar to Figure \ref{fig:CaseA_SS}, where the alternate masses are used (Case B + SS).  Even rotation periods beyond 24 hr are included to demonstrate the large scale variations of obliquity.}
\label{fig:CaseB_SS}
\end{figure}

\begin{figure}
\centering
\epsscale{1.0}
\includegraphics[width=\linewidth]{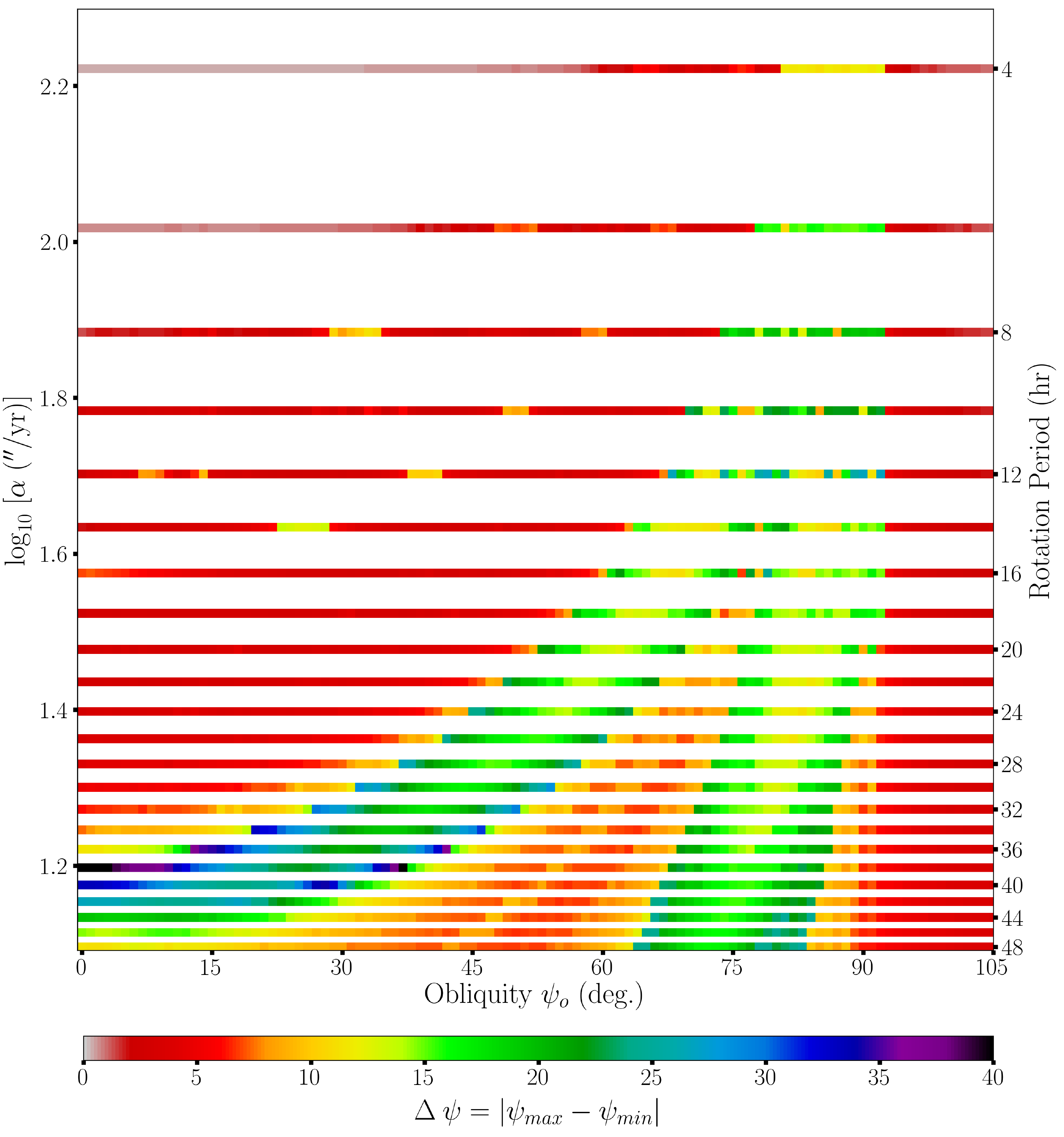}
\caption{Similar to Figure \ref{fig:CaseA_SS}, where the orbital inclinations of the Solar System giant planets are {doubled}.  The highest obliquity variations ($\psi_o < 5^\circ$ and P$_{rot} = 38$ hr) exceed the color scale with values of $\Delta\psi$ up to $42^\circ$.  The range of initial obliquity is expanded to 105$^\circ$ to show the transition from prograde to retrograde rotators, where the retrograde obliquities not shown ($\psi_o > 105^\circ$) have variations less than 5$^\circ$.  Even rotation periods are included to demonstrate the large scale variations of obliquity.  As a result, the maximum value of the color scale is changed.}
\label{fig:CaseA_SS_2x}
\end{figure}

\begin{figure}
\centering
\epsscale{1.0}
\includegraphics[width=\linewidth]{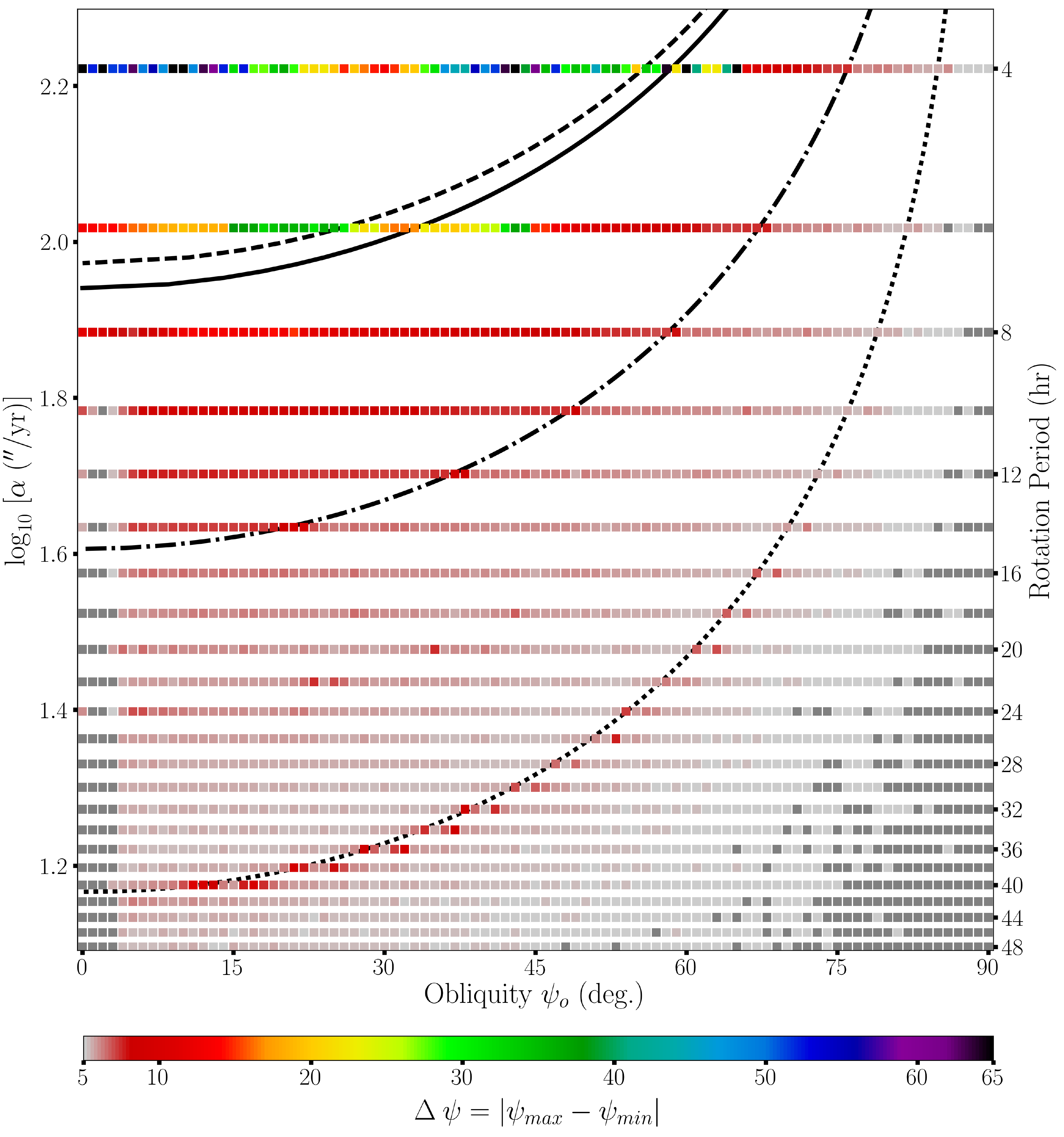}
\caption{Similar to Figure \ref{fig:CaseA_SS}, but considering our randomly determined Jupiter-Saturn pair (RG,see Section \ref{sec:outer_bodies}) instead (Case A + RG).  Even rotation periods are included to demonstrate the large scale variations of obliquity.  As a result, the minimum and maximum values of the color scale are changed, where obliquity variations below $5^\circ$ are all colored dark gray.  The black curves (solid, dashed, dotted, \& dash-dot) represent where the respective orbital precession frequencies ($f_2$, $f_4$, $f_5$, \& $f_8$) in Table \ref{tab:freq} overlap with the precession constant.}
\label{fig:CaseA_RG}
\end{figure}

\begin{figure}
\centering
\epsscale{1.0}
\includegraphics[width=\linewidth]{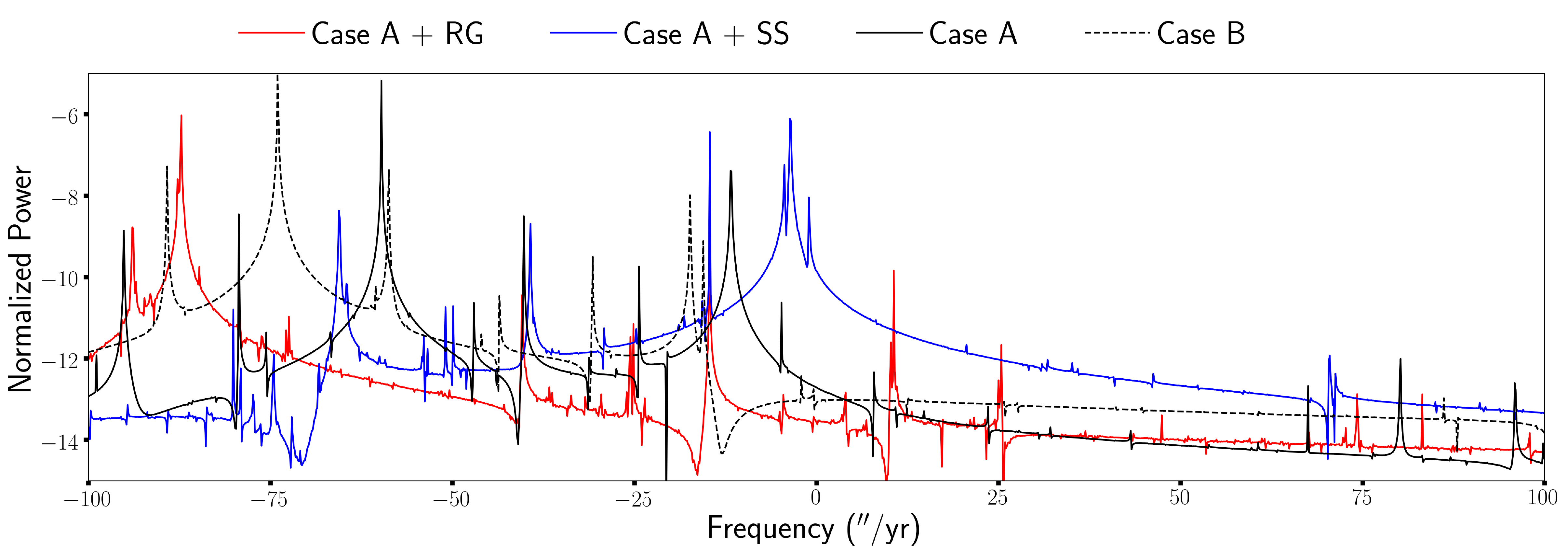}
\caption{Fourier spectra illustrating the relevant orbital frequencies (\arcsec/yr) using the inclination vector ($i\cos\:\Omega$, $i\sin\:\Omega$) of Kepler-62f while using the planet masses from Case A (black, solid) and Case B (black, dashed).  The Fourier spectra of Kepler-62f including the scaled Solar System giants (SS, blue) and our random gas giant pair (RG, red) are also shown.  The vertical axis is on a logarithmic scale.}
\label{fig:K62f_spectra}
\end{figure}

Our simulations (Figs. \ref{fig:CaseA} -- \ref{fig:CaseA_RG}) show that particular regions of parameter space are more likely to exhibit larger variations of obliquity, $\Delta\psi$, relative to other regions.  Figure \ref{fig:CaseA} indicates that 60 \arcsec/yr is a particularly important frequency given that the largest variation occurred for $\psi_0 = 0^\circ$, which corresponds to the the case $\dot{\Omega}_s = \alpha$ (Eq. \ref{eq:axial}).  In order to identify the frequencies more precisely, we apply a fast Fourier Transform (FFT) on the inclination vector of Kepler-62f using a 10 Myr dataset with samples every 1000 yr.  Figure \ref{fig:K62f_spectra} illustrates the resulting Fourier spectra considering four of our assumed architectures: (1) the 5 planet system with masses from Case A, (2) the 5 planet system with masses from Case B, (3) the 9 planet system with the Solar System giants (Case A + SS), and (4) the 7 planet system with our Jupiter-Saturn pair (Case A + RG).  There are several active peaks (power larger than --11, log scale) in each spectrum.  We note that the highest peak at 0 \arcsec/yr has been removed from each spectrum.

\begin{deluxetable}{c|ccc|ccc|ccc|ccc} 
\rotate
\tablecolumns{13}
\tablewidth{0pc}
\tabletypesize{\footnotesize}
\tablecaption{Top 10 Frequencies Determined Through FMFT Analysis\label{tab:freq}} 
\tablehead{   & \multicolumn{3}{c|}{Case A} & \multicolumn{3}{c|}{Case B} & \multicolumn{3}{c|}{Case A + SS} & \multicolumn{3}{c}{Case A + RG} \\
{$j$} & {$f_j$ (\arcsec/yr)} &  {$B$ (deg.)} & {$\gamma_j$ (deg.)} & {$f_j$ (\arcsec/yr)} &  {$B$ (deg.)} & {$\gamma_j$ (deg.)} & {$f_j$ (\arcsec/yr)} &  {$B$ (deg.)} & {$\gamma_j$ (deg.)} & {$f_j$ (\arcsec/yr)} &  {$B$ (deg.)} & {$\gamma_j$ (deg.)}}
\startdata
1	 &	0.00	&	0.41910	&	90.3	&	0.00	&	0.39692	&	90.3	&	0.00	&	1.57025	&	107.7	&	0.00	&	1.27783	&	91.6	 \\
2	&	-59.73	&	0.08776	&	93.0	&	-73.96	&	0.10279	&	90.6	&	-3.57	&	0.72488	&	283.4	&	-87.25	&	0.38749	&	260.5	 \\
3	&	-11.73	&	0.01468	&	271.7	&	-89.19	&	0.01024	&	251.1	&	-14.64	&	0.34227	&	293.0	&	-87.67	&	0.08507	&	30.1	 \\
4	&	-79.30	&	0.00331	&	255.4	&	-58.74	&	0.00981	&	110.0	&	-4.35	&	0.16848	&	316.5	&	-93.89	&	0.03370	&	231.5	 \\
5	&	-40.15	&	0.00326	&	110.7	&	-17.33	&	0.00530	&	284.2	&	-1.00	&	0.08441	&	22.6	&	-14.67	&	0.01218	&	270.2	 \\
6	&	-95.08	&	0.00248	&	352.4	&	-15.52	&	0.00165	&	232.0	&	-65.51	&	0.07116	&	82.8	&	-86.82	&	0.01176	&	129.3	 \\
7	&	-24.37	&	0.00085	&	13.6	&	-30.75	&	0.00105	&	32.5	&	-39.30	&	0.04123	&	126.8	&	10.62	&	0.00746	&	211.1	 \\
8	&	-47.08	&	0.00034	&	351.1	&	-43.51	&	0.00052	&	129.4	&	-64.47	&	0.01153	&	170.7	&	-40.39	&	0.00516	&	38.9	 \\
9	&	-4.80	&	0.00029	&	211.3	&	-60.54	&	0.00020	&	342.4	&	-4.04	&	0.00606	&	232.9	&	-91.40	&	0.00469	&	225.1	 \\
10	&	-59.04	&	0.00016	&	284.7	&	-75.77	&	0.00017	&	323.5	&	-50.96	&	0.00467	&	107.9	&	-90.97	&	0.00377	&	275.0 
\enddata
\tablecomments{Top 10 values of the frequency ($f_j$), amplitude ($B$), and the phase ($\gamma_j$) found using the FMFT analysis for our 5 planet systems (Case A \& Case B), as well as for our simulations that include the scaled Solar System giants (Case A + SS) or a Jupiter-Saturn pair of gas giants (Case A + RG).} 
\end{deluxetable}

Table \ref{tab:freq} shows the top 10 frequencies ($f_j$) identified in the time series using Frequency Modified Fourier Transform\footnote{\url{https://www.boulder.swri.edu/~davidn/fmft/fmft.html}} \citep{Sidlichovsky1996}.  The rows of Table \ref{tab:freq} are ordered by the amplitude of each mode, where the index $j$ refers to the counting of the modes and does not correspond to a particular body.  We list for each scenario the frequency ($f_j$ in \arcsec/yr), the amplitude ($B$ in degrees), and the phase of the mode ($\gamma_j$ in degrees).  The highest amplitude frequency ($j_1$) occurs at 0 \arcsec/yr, which arises from a degeneracy in inclination vectors \citep{Murray2000}.  \cite{Shan2018} performed an analytical analysis of Kepler-62f using the Lagrange-Laplace method and found similar frequencies present.  Also, the instantaneous precession period can be determined using these frequencies, where the high frequency terms produce precession periods much faster than the present day Earth.

Figure \ref{fig:CaseA} shows large variations near 60 \arcsec/yr, and from Table \ref{tab:freq} we expect this to occur most strongly at 59.73 \arcsec/yr.  Another region of obliquity variation, although much smaller, occurs at $\alpha \approx 25$ \arcsec/yr and $\psi_o \approx 62^\circ$.  At this location, we find that $\dot{\Omega}_s = 11.7$ \arcsec/yr, which is approximately equal to the $f_3$ frequency for Case A.  In Section \ref{sec:5pl_var}, we found that the region of largest variation changed for Case B (Fig. \ref{fig:CaseB}) shifting to higher frequencies, {which is because Kepler-62e has a larger assumed mass and induces a higher induced precession frequency}.  When looking at the Fourier spectra (Fig. \ref{fig:K62f_spectra}) and the associated frequencies (Table \ref{tab:freq}), it is apparent that a shift to higher frequencies has occurred and explains the large scale differences between our results in Figs. \ref{fig:CaseA} and \ref{fig:CaseB}.

Adding the Solar System giants to Case A (Case A + SS) produces much larger variations in obliquity (Fig. \ref{fig:CaseA_SS}) in the low frequency regime.  The region of largest variation (long rotation period, low initial obliquity) can be associated with the $f_3$ frequency, where the high obliquity regions is associated with the $f_2$ frequency.  The associated frequencies for the scenario where we double the Solar System giants' inclination is very similar, where the amplitudes ($B$) are approximately double.  However, $f_{10}$ is different with values of --2.57 \arcsec/yr,  0.00824$^\circ$, and 70.0$^\circ$ for $f_{10}$, $B$, and $\gamma_{10}$, respectively.  This distinction is important because Fig. \ref{fig:CaseA_SS_2x} shows 2 regions for P$_{rot}$ = 38 hr with $\Delta\psi \approx 42^\circ$, which corresponds to the $f_3 - f_5$ ($\psi_o \approx 0^\circ$) and $f_3 - f_{10}$ ($\psi_o \approx 40^\circ$) frequency combinations.  Our Jupiter-Saturn pair (Case A + RG) is dominated by much higher frequencies and even includes a positive frequency (10.62 \arcsec/yr) that would make a backwards rotating (retrograde) Kepler-62f interesting.

\begin{figure}
\centering
\epsscale{1.0}
\includegraphics[width=\linewidth]{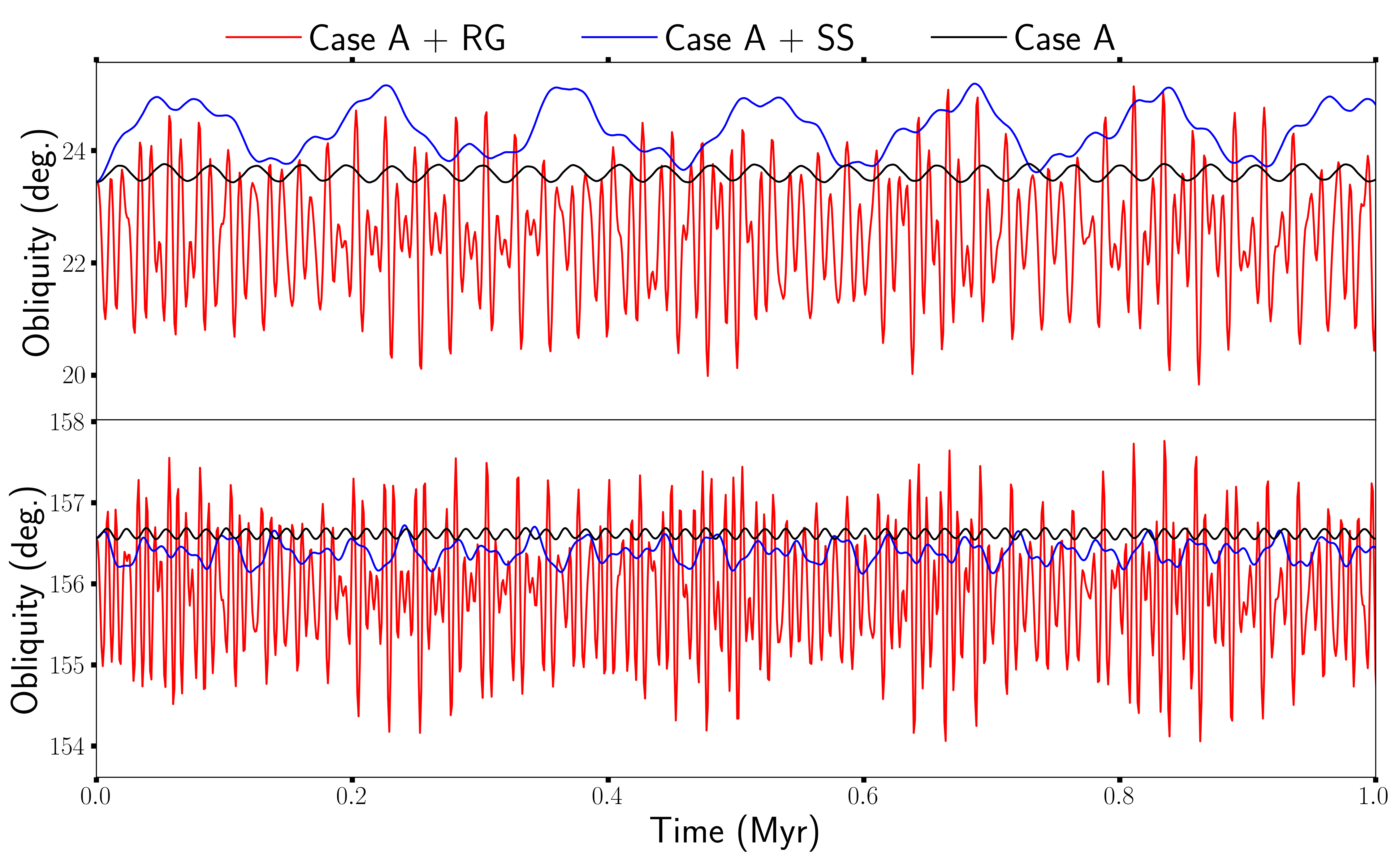}
\caption{Evolution of a prograde (top) and retrograde (bottom) Kepler-62f with Earthlike values in initial obliquity and rotation period while using the planet masses from Case A (black).  The evolution of these parameters including the scaled Solar System giants (SS, blue) and our random gas giant pair (RG, red) are also shown.}
\label{fig:K62f_obl}
\end{figure}

The obliquity evolution over the first 1 Myr for a Earthlike ($\psi_o = 23.44^\circ$ \& P$_{rot} = 23.934$ hr) Kepler-62f is shown in Figure \ref{fig:K62f_obl} for both a prograde (top) and retrograde (bottom) rotator.  The evolution of the 5 planet system (Case A, black) displays very small variations, where those with the Solar System giants added (Case A + SS, blue) are more substantial.  The evolution, when including our Jupiter-Saturn pair, is much larger ($\sim$5$^\circ$) in both prograde and retrograde.  We note that the frequency of variation between the prograde and retrograde rotators (Case A + RG, red) is slightly different, where this is due to the overlap of slightly different positive and negative orbital frequencies ($f_5$ \& $f_7$, Table \ref{tab:freq}).

The orbital obliquity can change depending on the assumed masses (Case A or Case B), the outer bodies (possible giant planets), and the assumed longitude of the spin node ($\Omega_s$), where $\Delta\Omega = \Omega_s - \Omega$, the difference between the spin node and the orbital node, is the more important quantity.  In almost all of our simulations, we have assumed that $\Delta\Omega = 0^\circ$ based upon our previous results in \cite{Barnes2016}.  Our previous finding showed little change because the phase of the secular frequencies ($\gamma_j$) in the Solar System were not near the values of $\Delta\Omega$ that we simulated.  From Table \ref{tab:freq}, we can see that the strongest non-zero frequency ($j_2$) has a phase angle near 90$^\circ$.

\begin{figure}
\centering
\epsscale{1.0}
\includegraphics[width=\linewidth]{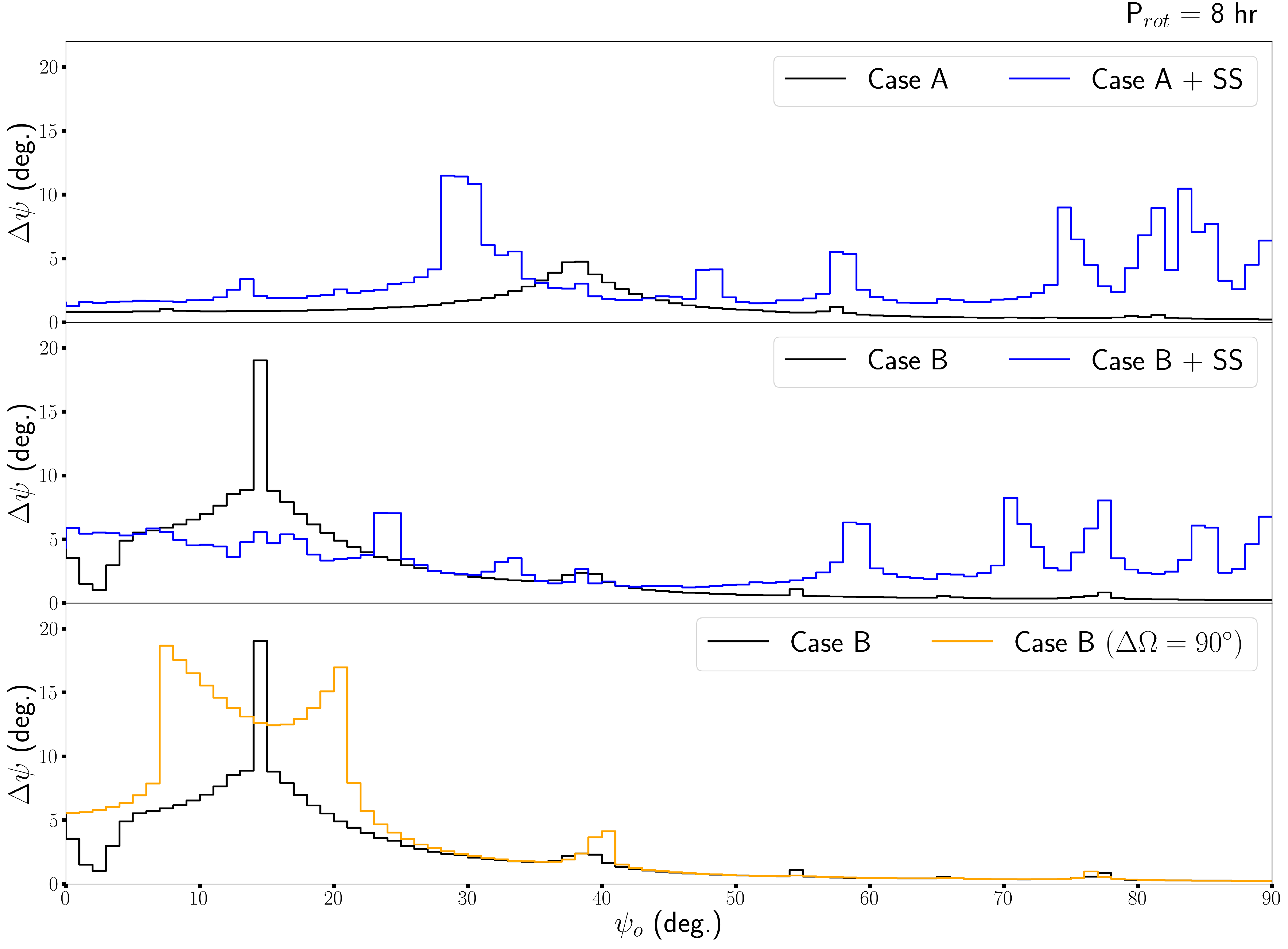}
\caption{Variation in obliquity, $\Delta\psi$, for Case A and Case B (black) as a function of the initial obliquity for a rotation period of 8 hours.  Variations when adding the Solar System (SS) giants are also shown in blue.  The bottom panel shows the same results from the middle panel (Case B), but adds results of simulations that modify the relative nodal angle by 90$^\circ$ (orange). }
\label{fig:var}
\end{figure}

Figure \ref{fig:var} shows the variation in obliquity ($\Delta\psi$) considering a prograde Kepler-62f with an 8 hr rotation period.  The top and middle panels demonstrate the difference in variation between the 5 planet systems (Case A \& Case B) and their respective 9 planet systems where the Solar System (SS) giant planets are added. Similar peaks are present that correspond to color changes in Figs. \ref{fig:CaseA} -- \ref{fig:CaseB_SS} for an 8 hr rotation period.  The bottom panel of Fig. \ref{fig:var} shows the changes due to our assumption on $\Delta\Omega$.  The region of significant obliquity variations occurs near the same initial obliquity ($\psi\sim 16^\circ$), but the range of initial obliquity values, $\psi_o$, is broadened when we consider $\Delta\Omega = 90^\circ$.  Additionally, the variation near $\psi_o = 40^\circ$ is larger.  

\subsection{Effects on the Potential Habitability}\label{sec:hab}
{The average flux received by the Earth is larger than what Kepler-62f ($\sim60\%$  less) receives due to differences in orbital distance relative to the difference between the host stars (see Equation \ref{eq:flux}).  Thus, if Kepler-62f is habitable, irrespective of its changes in obliquity, then its atmosphere must be different such that a more significant greenhouse effect is present.  \cite{Shields2016} showed, using 3D Global Circulation Models, that a CO$_2$ dominated atmosphere with 5 bars of atmospheric pressure would allow Kepler-62f to be considered habitable by current standards.}  Several studies of the habitability of a planet include many such assumptions that vary between models \citep[e.g.,][]{Kasting1993,Kopparapu2013,Kopparapu2014,Kopparapu2016}, where we present results that survey the effect of coupled orbital and obliquity variation on the latitudinal surface flux \citep[e.g.,][]{Williams2002,Williams2003,Armstrong2004,Armstrong2014,Shields2016,Kane2017,Kilic2017}.  {We focus on the effects of obliquity variation relative to what the modern Earth experiences using our numerical simulations (Case A, Case A + SS, Case A + RG) including the some of the resonant cases that induce large obliquity variations.  Some of these cases occur for faster rotation rates than Earth, which could be important because others have suggested that faster rotation rates can increase the prospects of habitability for Kepler-62f \citep{Ramirez2018}.}

\begin{figure}
\centering
\plotone{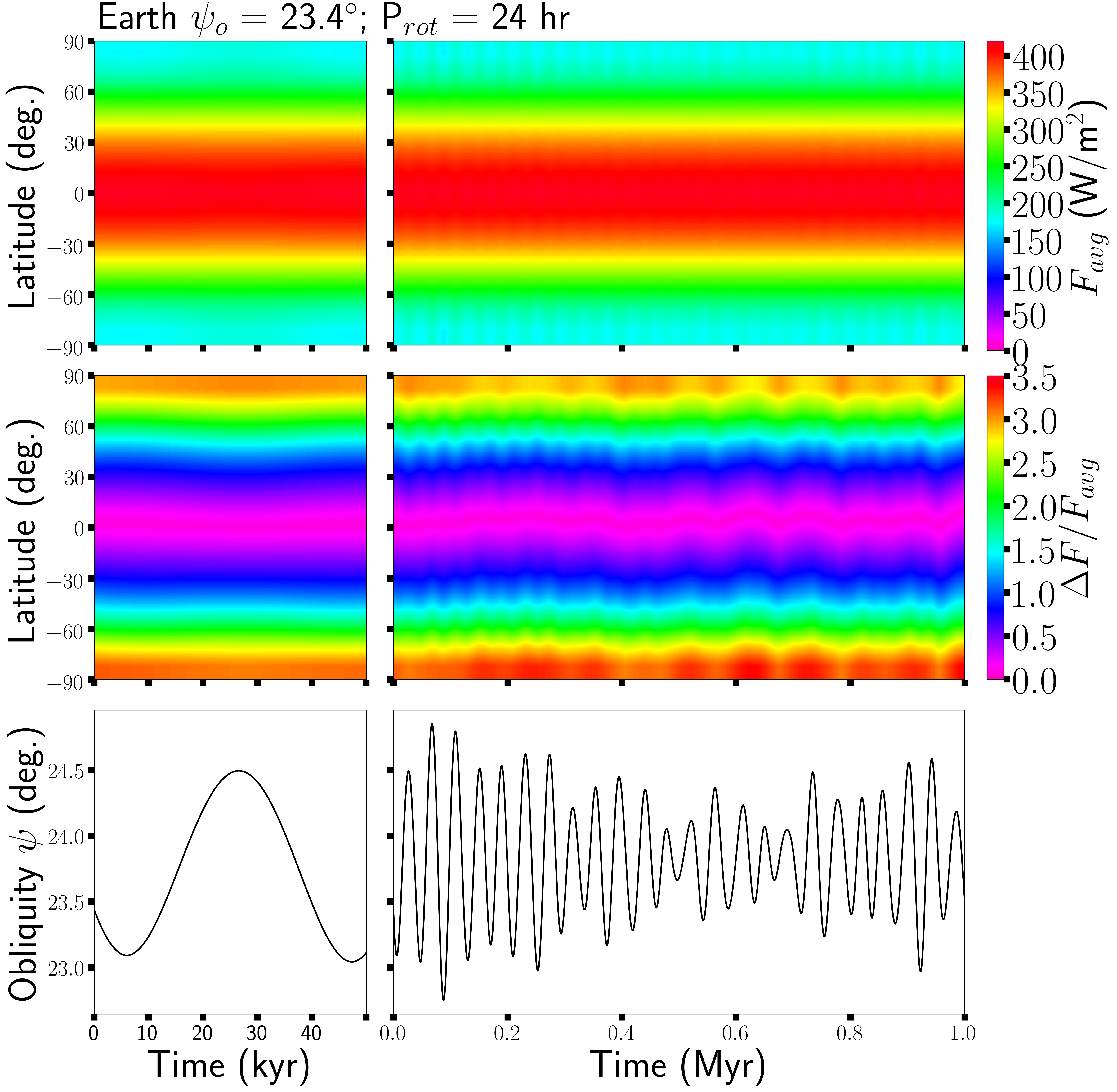}
\caption{Latitudinal surface flux variations of the Earth as a result of obliquity variations over short (50 kyr) and long (1 Myr) timescales.  The top row illustrates the mean annual flux ($F_{avg}$) as a function of time, the middle row identifies the relative change in flux ($\Delta F = F_{max} - F_{min}$) over an orbit, and the bottom row shows the evolution of obliquity for the respective timescales. }
\label{fig:Earth_flux}
\end{figure}

{First, we examine the flux variations of the modern Earth so that our later results can be contrasted and placed into context.  The mean annual flux $F_{avg}$, as shown in Figure \ref{fig:Earth_flux} (top row), appears largely stratified where the equator receives the bulk ($\sim400$ W/m$^2$) of the radiation and the poles receive substantially less radiation from the Sun ($\sim175$ W/m$^2$).  Although the mean annual flux appears roughly constant, the latitudinal flux changes over a yearly cycle (Fig. \ref{fig:Earth_flux}; middle row), where the polar regions can experience the most dramatic effects with differences of $\sim$3.5x the mean flux between the summer and winter extremes.  Our simulations reproduce the expected obliquity variation of the modern Earth ($\pm 1.3^\circ$ over 41,000 years) which causes regular climatic shifts (Fig. \ref{fig:Earth_flux}; bottom row)).  In addition, the non-periodic shifts in the obliquity are seen the in the fractional change of the flux ($\Delta F$/$F_{avg}$).}

\subsubsection{Earthlike and Resonant Spins of Kepler-62f} \label{sec:K62f_flux}
{As noted before, Kepler-62f resides in a relatively more distant orbit than the Earth and we expect the magnitude of the mean annual flux to differ.  Figure \ref{fig:K62f_flux} illustrates Earthlike conditions in terms of the spin state ($\psi_o = 23.4^\circ$, P$_{rot} = 24$ hr), where the mean annual flux at the equator ($\sim170$ W/m$^2$) more closely resembles Earth's polar regions (top row).  Apart from the difference in magnitude, Figs. \ref{fig:Earth_flux} and \ref{fig:K62f_flux} (top rows) appear quite similar in structure.  Differences appear to arise when we consider the fractional change (Figs. \ref{fig:Earth_flux} and \ref{fig:K62f_flux} (middle rows)) of the flux  in the southern polar region ($\sim$4.25x compared to $\sim$3.5x), but the absolute differences $\Delta F$ ($\sim$300 W/m$^2$ compared to $\sim$590 W/m$^2$ ) shows that the changes between summer and winter can be milder than those experience at Earth's south pole due to the lower mean annual flux $F_{avg}$.  However the minimum flux at the poles for Kepler-62f are much lower than what the Earth experiences.  For the obliquity evolution (\ref{fig:K62f_flux}, bottom row), we find the precession period to be similar to the Earth, but with more periodic variations.  The similarity of precession period of the Earth is purely coincidental with the spin precession period for this test case.  The more periodic nature of the obliquity evolution, on the other hand, comes from the relatively weak perturbations of the neighboring planets in Kepler-62.}

{Figure \ref{fig:CaseA} illustrates the locations where spin-orbit coupling can play a significant role and thereby induce large obliquity variations.  We examine, in Figure \ref{fig:K62f_res_flux}, the flux variations when Kepler-62f begins with a shorter rotation period (10 hr) and near a commensurability with $f_2$ (see Table \ref{tab:freq}).  The relatively short term ($<50$ kyr) evolution shows the expected result for $\psi \approx 0^\circ$ where the mean annual flux $F_{avg}$ to be extremely stratified and the poles receive negligible amounts of radiation even when accounting for the yearly variation $\Delta F$.  The obliquity slowly increases on a $\sim$1 Myr timescale to $\sim$20$^\circ$ from to the resonant effect of the spin-orbit interactions, which causes the mean annual flux to increase at the poles up to a maximum ($\sim$70 W/m$^2$) and the difference between summer and winter extremes is $\sim$210 -- 280 W/m$^2$.  Figure \ref{fig:K62f_res_flux} (middle rows) shows a ringing effect, which illustrates the effects on flux the variations due to eccentricity variations (inset panel).  Flux variations due eccentricity are also apparent in the previously discussed Earthlike rotator case (Figure \ref{fig:K62f_flux}; middle row).}

\subsubsection{Effects of Giant Planets on Kepler-62f}
{From Section \ref{sec:outer_bodies}, we demonstrate that the addition of giant planets on longer period orbits increases obliquity variation across the parameter space and introduces new regions at longer rotation periods where large obliquity variation is possible.  Here we examine where the spin-orbit interactions produce significant obliquity variation including: (1) the Solar System giants (scaled by period to Kepler-62) and (2) a random Jupiter-Saturn pair.}

{Figure \ref{fig:K62f_resSS_flux} considers a slowly rotating planet (P$_{rot} = 40$ hr) with a nearly Earthlike initial obliquity ($\psi_o = 28^\circ$).   When the scaled Solar System giants are included the mean annual flux (top rows) is initially similar in structure to the Earthlike case discussed in Section \ref{sec:K62f_flux}, but the poles receive slightly more average radiation ($\sim$100 W/m$^2$) due to the increased obliquity.  Over much longer timescales (1.5 Myr) the obliquity decreases to nearly $0^\circ$ and the mean annual flux changes dramatically.  This becomes important to habitability. In our solar system, for instance, long periods at low obliquity contributed to the collapse of Mars’ atmosphere assuming it began with a more substantial CO$_2$
atmosphere \citep{Forget2013}.  There are variations in the seasonal extremes in flux ($\sim$3.5 - 4.25x $F_{avg}$) at the poles (Figure \ref{fig:K62f_resSS_flux}, middle rows) that comes from the eccentricity variation.  Although the fractional change remains high, the magnitude of the change decreases as the obliquity decreases. The obliquity changes by $\sim$ 26$^\circ$ over a 1.5 Myr timescale, where the effects of this change (in terms of the flux) can be quite dramatic when including those due to the eccentricity.}

{We examine another giant planet case but including the Jupiter-Saturn pair (Case A + RG) to see how the obliquity of a rapid rotator ($P_{rot}$ = 4 hr) evolves and the impact on the flux variations.  Figure \ref{fig:K62f_resRG_flux} shows that over a 10 Myr timescale the obliquity changes (bottom rows) greatly affect the mean annual flux (top rows) and the fractional change in flux (middle rows).  Initially the mean annual flux is similar to the previous cases, but as the obliquity increases (over the first 50 kyr) the polar regions receive more radiation per year.  After $\sim$3.5 Myr there is a shift towards even higher obliquity, but the year-to-year variation is less.  Another shift occurs at $\sim$7.5 Myr that pushes the planet into a high obliquity regime, where the mean annual flux at the poles is relatively high and there are large seasonal variations ($\Delta F \gtrsim 400$ W/m$^2$).  The evolution of these states occur over millions of years, but there are times of stark transition that may be detrimental in the current view of habitability, if such conditions exist.}

\section{Conclusions}
We explore possible dynamical states of the Kepler-62 system focusing on the variation of obliquity for the outermost planet, Kepler-62f, due to its high astrobiological interest \citep{Borucki2013,Ramirez2018}.  The possible obliquity variations of Kepler-62f depend on many unknowns including the masses of the five known planets (Case A or Case B), the possible presence of outer bodies (Case A/B + SS or Case A + RG), the direction of the spin, and the relative nodal angle, $\Delta\Omega$.  Each of these assumptions can introduce variations larger than the present day Earth (including the stabilizing effects of the Moon) on timescales of a few million years, where the largest contributor is the presence of gas giants with larger orbital periods and inclinations that could have escaped detection.  We determine the magnitude of these variations using N-body simulations and their relation to the assumed rotation parameters using frequency analysis.  {The flux received at the top-of-the atmosphere is measured on both short (50 kyr) and long (1 Myr) timescales for a range of representative cases, where the effects of planetary eccentricity and obliquity evolution are present.}

\begin{figure}
\centering
\plotone{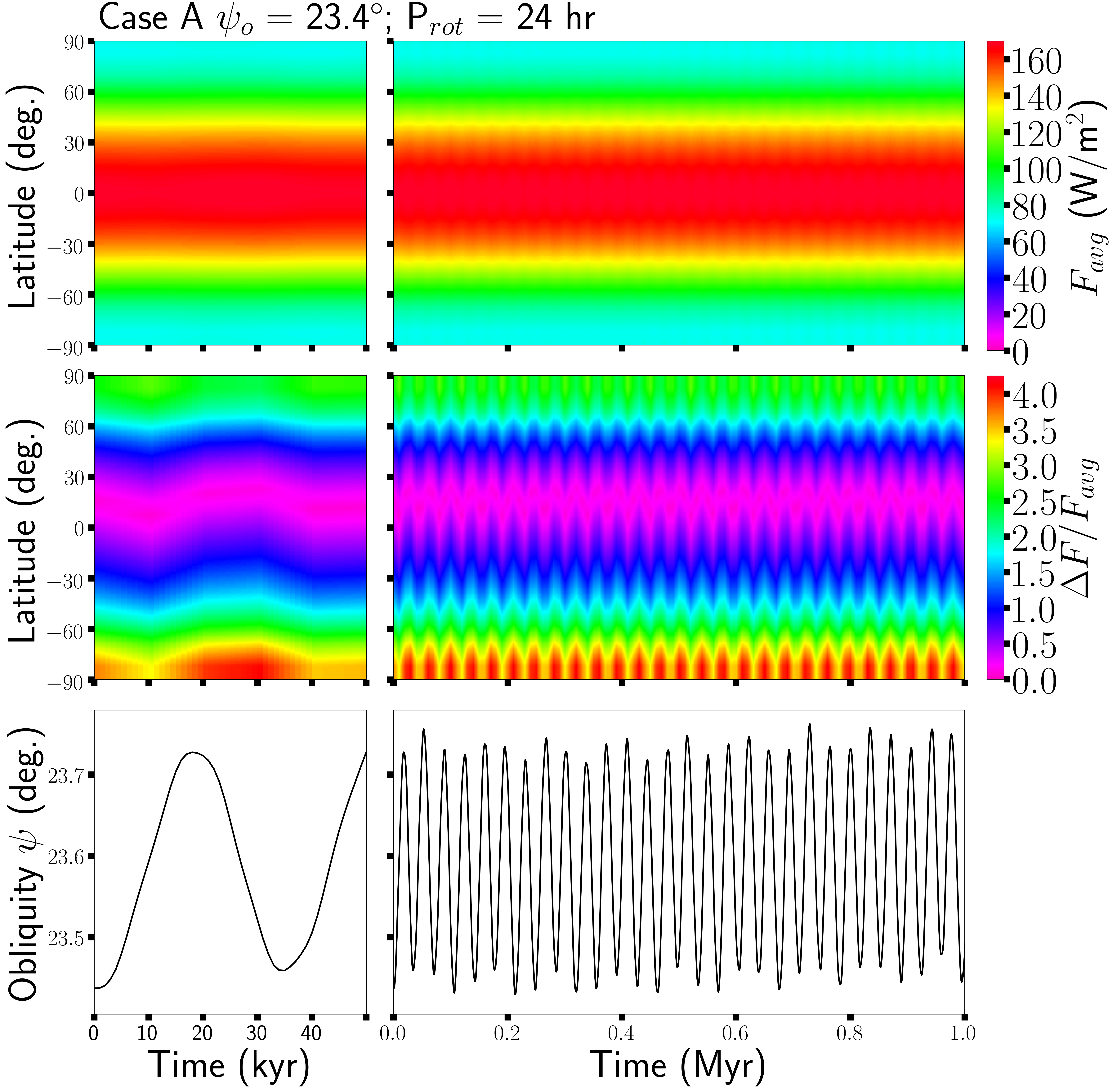}
\caption{Similar to Figure \ref{fig:Earth_flux}, but considering Kepler-62f (using Case A masses) with an Earthlike initial spin state ($\psi_o = 23.4^\circ$, P$_{rot}$ = 24 hr).  We note that the scale of the color-code changes in this figure and subsequent figures.}
\label{fig:K62f_flux}
\end{figure}

\begin{figure}
\centering
\plotone{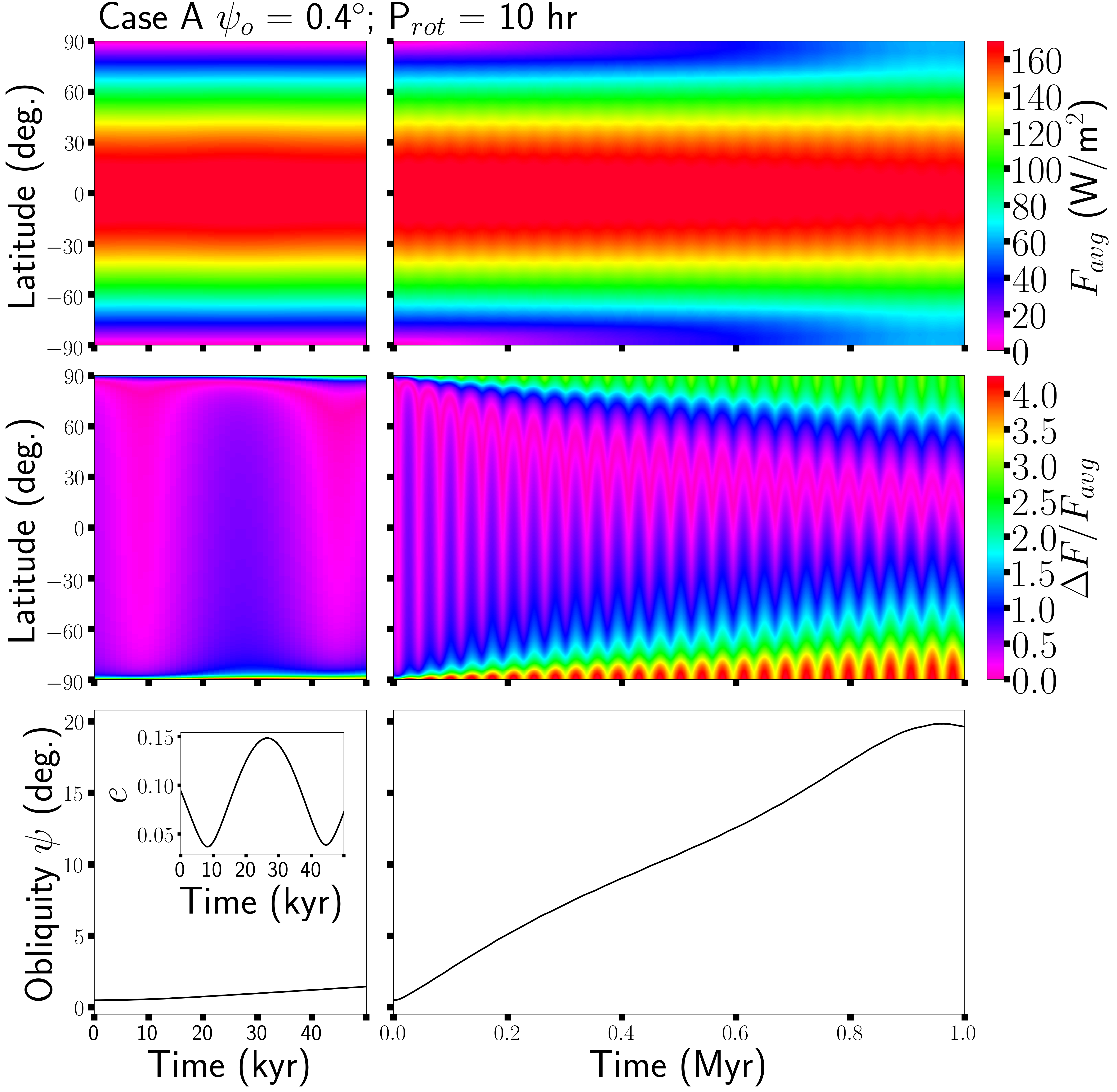}
\caption{Similar to Figure \ref{fig:K62f_flux}, but considering Kepler-62f (using Case A masses) with an initial spin state near the spin orbit resonance ($\psi_o = 0.4^\circ$, P$_{rot}$ = 10 hr).}
\label{fig:K62f_res_flux}
\end{figure}

\begin{figure}
\centering
\plotone{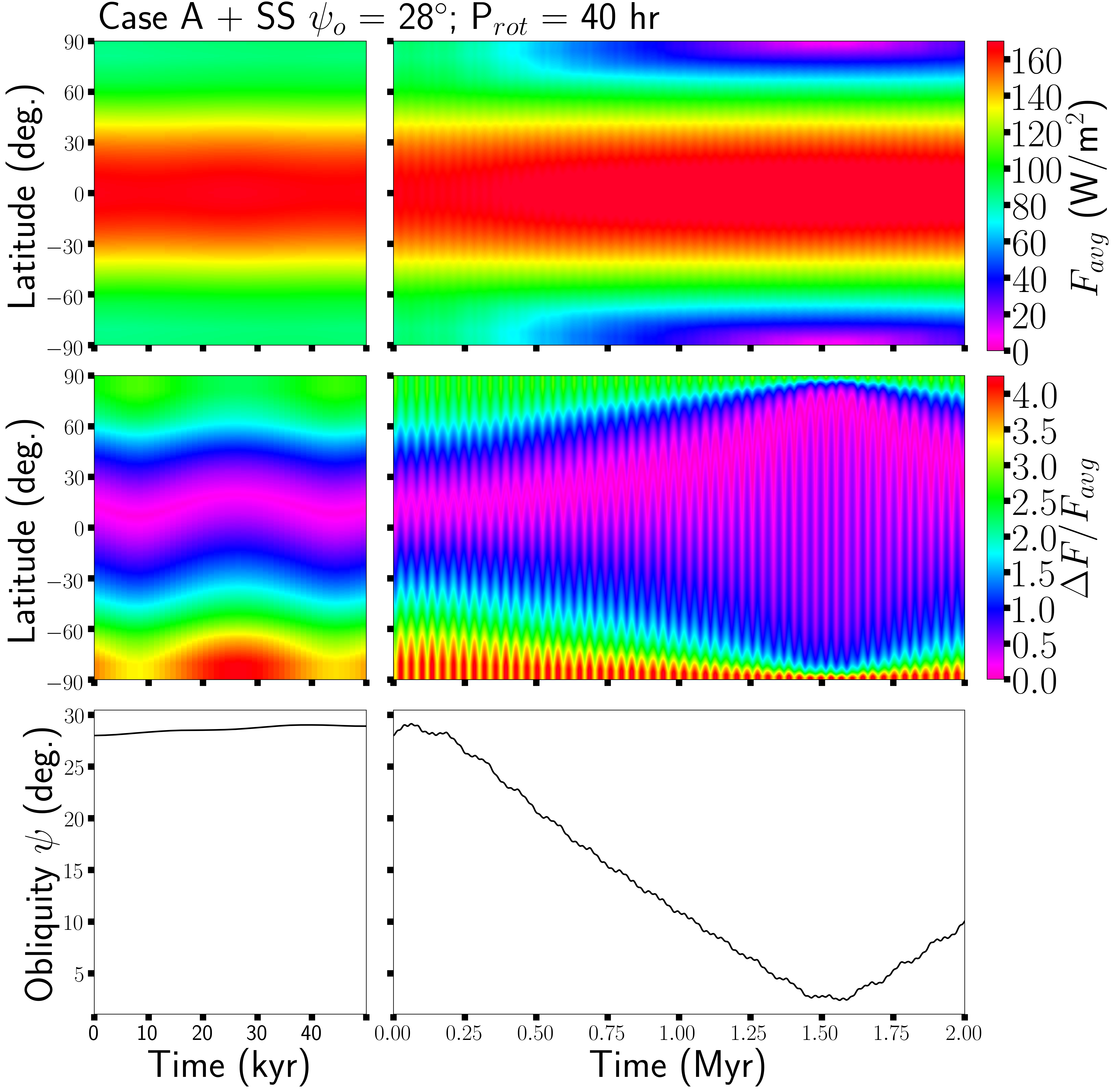}
\caption{Similar to Figure \ref{fig:K62f_res_flux}, but considering Kepler-62f (using Case A + SS masses) with an initial spin state near the spin orbit resonance ($\psi_o = 15^\circ$, P$_{rot}$ = 15 hr).}
\label{fig:K62f_resSS_flux}
\end{figure}

Regions of significant variation in obliquity differ based upon our initial assumptions of the masses (and compositions) of the Kepler-62 planets.  Considering a 0$^\circ$ initial obliquity and a 10 hr rotation period for Kepler-62f (Case A) produces a $\sim$20$^\circ$ difference between the highest and lowest attained obliquity, while most other choices for initial obliquity and rotation period are limited to variations less than $3^\circ$.  {We show the latitudinal surface flux to vary in response to the orbital and obliquity evolution of the system (see Fig. \ref{fig:K62f_flux}) resulting in a variation of $\sim$280 W/m$^2$ in the mean annual surface flux at the poles.}  A similar range of variation in Case B occurs at a higher initial obliquity ($\psi_o \sim 17^\circ$) when considering a different mass-radius relation for the inner five planets, but requires a faster rotation (8 hr period) due to an induced precession from Kepler-62e.  The range in initial obliquity that produces moderate variations ($>10^\circ$) can broaden due to overlap between a secular mode and the relative nodal longitude, $\Delta\Omega$.

Including a set of giant planets similar to Solar System to either case (Case A or Case B) increases the overall obliquity variation, but not substantially for most cases.  In order for strong variations ($>25^\circ$) to occur, the rotation period of Kepler-62f needs to be longer than 24 hr and this also depends on the initial obliquity.  Solar System-like giant planets with double their orbital inclinations produce much broader regions of moderate obliquity variation, where strong variations ($\Delta\psi \approx 42^\circ$) can occur for specific initial parameters.  Retrograde obliquities ($\psi_o > 90^\circ$) for the double inclination scenario (which exhibits the largest variations for large prograde obliquities) are large for initial obliquity in the range $95^\circ > \psi_o > 90^\circ$ , but typically produce relatively low obliquity variations ($\Delta\psi < 5^\circ$) for obliquity above $95^\circ$.

\begin{figure}
\centering
\plotone{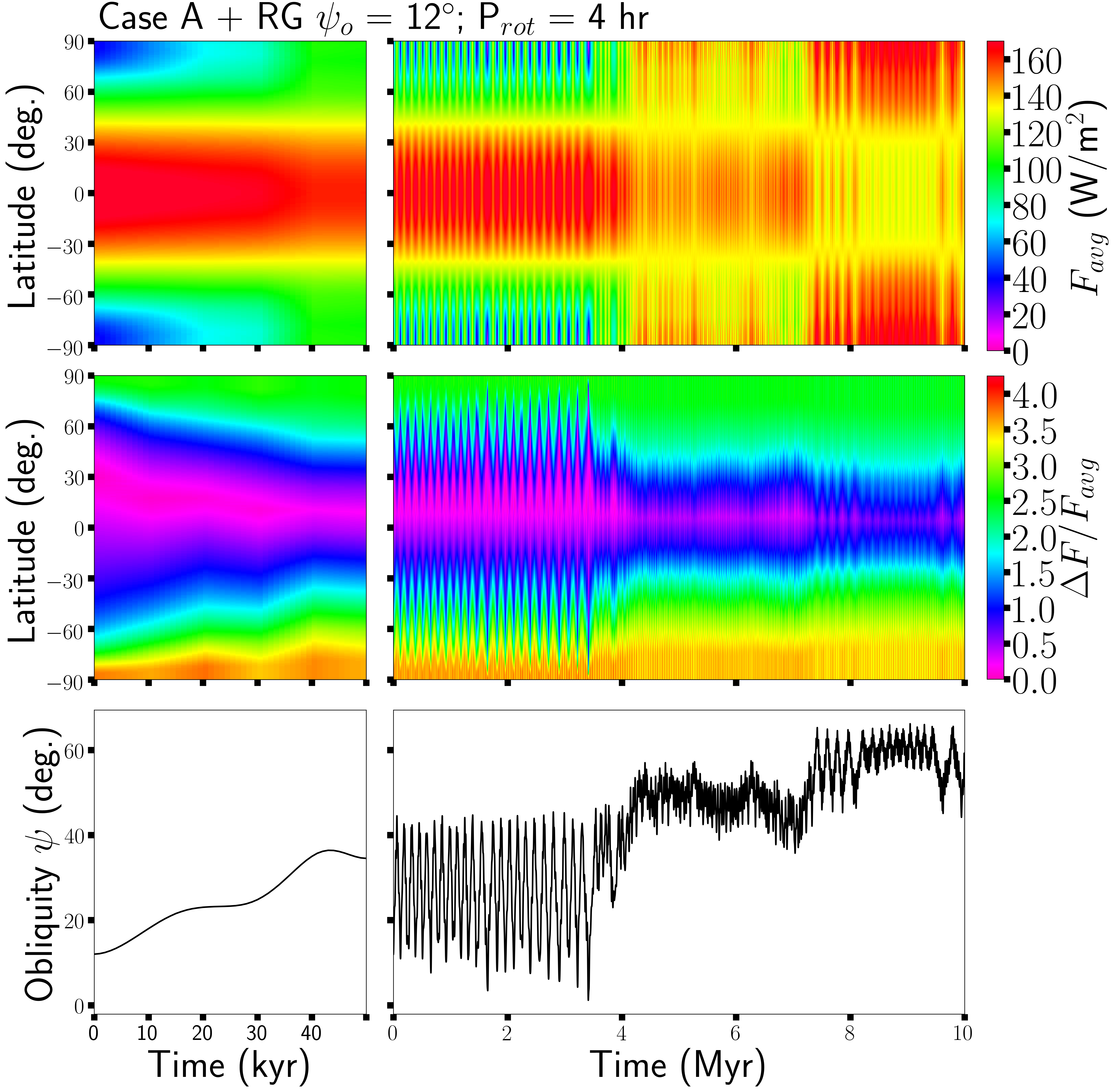}
\caption{Similar to Figure \ref{fig:K62f_res_flux}, but considering Kepler-62f (using Case A + RG masses) with an initial spin state near the spin orbit resonance ($\psi_o = 12^\circ$, P$_{rot}$ = 4 hr).}
\label{fig:K62f_resRG_flux}
\end{figure}

{Our 7 planet systems that include a pair of gas giants on close-in (1.8 -- 3.5 AU) orbits could induce high variations of obliquity ($\sim$66$^\circ$), but this requires Kepler-62f to be a relatively rapid ($<8$ hr) rotator {due to the higher  orbital precession frequency of the giant planets.}  The most common obliquity variation is larger than the present day Earth-Moon system, but not extremely high ($\Delta\psi < 10 ^\circ$).  We show the latitudinal surface flux to vary in response to the orbital and obliquity evolution of system with a 4 hour rotation period (see Fig. \ref{fig:K62f_resRG_flux}) resulting in a variation of $\sim$50-400 W/m$^2$ in the mean annual surface flux at the poles.  During epochs of high obliquity, the polar regions can receive a substantial surface flux ($\sim$400 W/m$^2$) at the poles for nearly a third of an orbit.  The obliquity can transition into different ranges on a 10 Myr timescale and thus dramatically affect the prospects of habitability. }

Obliquity variation can have an impact on the potential climates of exoplanets \citep{Williams1997,Spiegel2009}, where some climate model calculations have been performed specifically for the Kepler-62 system \citep{Bolmont2015,Shields2016}.  Although we do not include a full climate model in our analysis, we find that the obliquity evolution can differ substantially when additional planets on long period orbits are considered and thereby alter the amount of latitudinal surface flux that a planet receives at various epochs.  The amount of obliquity variation, $\Delta\psi$, can increase substantially and potentially affect the broader conclusions drawn about climates on potentially habitable worlds.  Recently, \cite{Ramirez2018} found that Kepler-62f is one of three confirmed exoplanets that lie within a zone of habitability for water worlds called the ice cap zone and a fast rotation rate ($\lesssim 8$ hours) would be necessary {to allow for habitability by most definitions}. 

{Our study probes the Kepler-62 system using the best estimates for the planetary masses and best known observationally derived orbital elements.  However, there remains significant uncertainty in these values when compared with those of the Solar System planets, which highlights the need for additional observations that could better constrain the system architecture for more robust studies.  Upcoming planet surveys (TESS, \cite{Ricker2014}) have prioritized searching for Earth-mass planets orbiting M dwarfs and it may be some time before another system dynamically similar to Kepler-62 (with a habitable zone planet) is discovered.  \cite{Kane2016} produced a categorized catalog of potentially habitable planets, where Kepler-62f is included in all four categories and indicates that it would be an ideal candidate for any observational follow-up program that targets habitable zone exoplanets.  }

\acknowledgments
The authors would like to thank the anonymous reviewers for constructive insights that greatly enhanced the quality and clarity of the manuscript.  In addition, the authors thank Gongjie Li for helpful discussions.  We are also grateful to Yutong Shan and Gongjie Li for providing detailed feedback on our draft.  The authors acknowledge support from the NASA Exobiology Program, grant \#NNX14AK31G.

\bibliographystyle{aasjournal}
\bibliography{bibliography}

\begin{figure}
\centering
\epsscale{1.0}
\includegraphics[width=\linewidth]{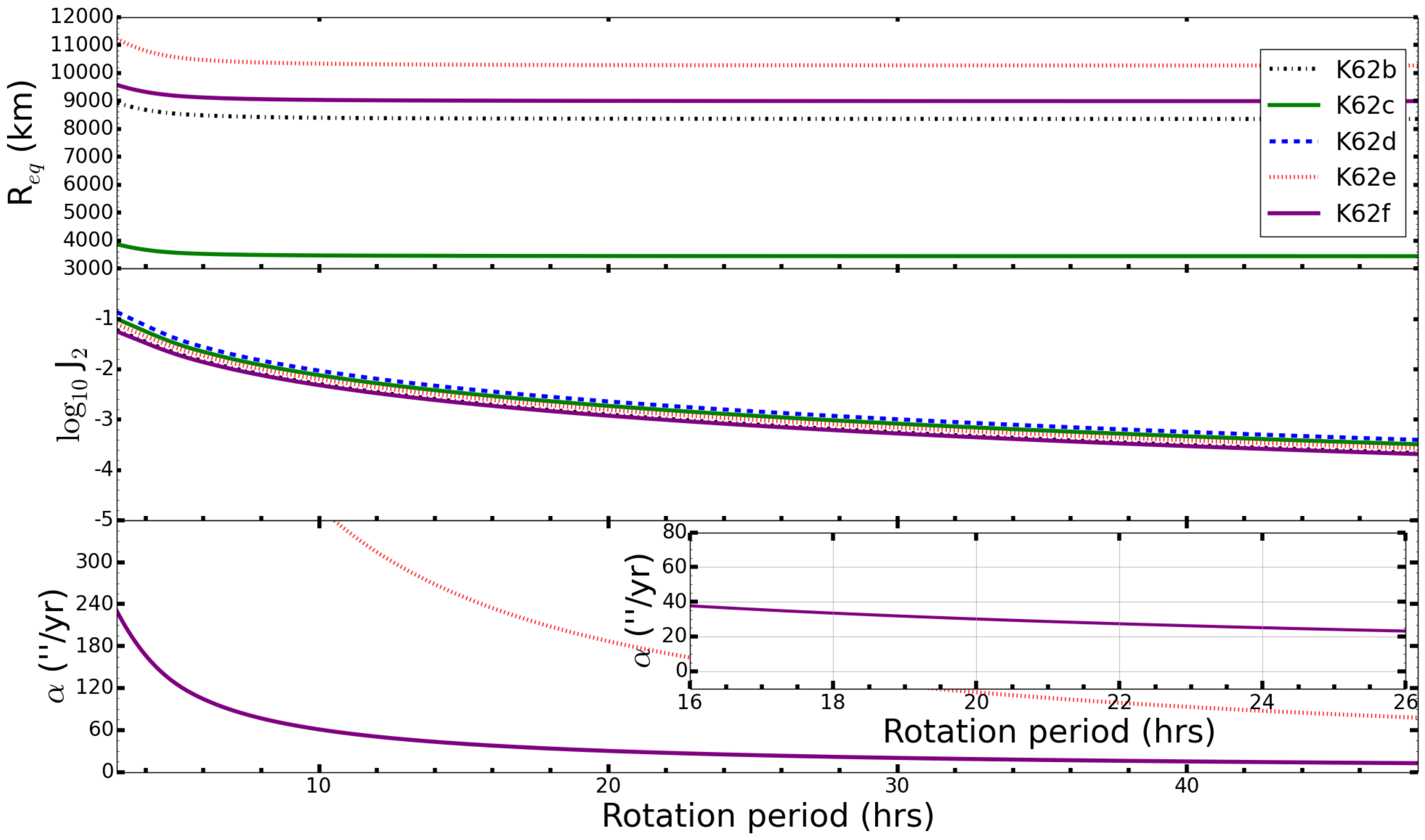}
\caption{Rotational parameters (equatorial radius, $J_2$, $\alpha$) using the masses in Case A for the Kepler-62 planets as a function of an assumed rotation period in hours.  The inset panel shows a zoomed view of Kepler-62f over a range of rotations periods similar to Earth.}
\label{fig:spin_param}
\end{figure}

\end{document}